\documentclass[aps,pra,a4paper,reprint,superscriptaddress]{revtex4-2}

\usepackage{float}
\usepackage[utf8]{inputenc}
\usepackage[T1]{fontenc}

\usepackage{graphicx}
\graphicspath{{figures/}}
\usepackage{calc}
\usepackage{amsmath,amssymb,amsthm}
\usepackage{mathtools}
\usepackage{braket}
\usepackage[dvipsnames]{xcolor}
\usepackage{graphicx}
\usepackage{booktabs}
\usepackage{enumerate}
\usepackage{setspace}
\usepackage{xr}
\usepackage{subfigure}
\usepackage{palatino}
\usepackage{times}
\usepackage{soul}
\usepackage{physics}
\usepackage[normalem]{ulem}

\usepackage[ruled,vlined]{algorithm2e}
\SetAlgorithmName{Procedure}{Procedure List}{List of Procedures}

\definecolor{pgreen}{RGB}{67, 143, 100}
\definecolor{porange}{RGB}{199, 103, 42}
\usepackage[
colorlinks,
citecolor=pgreen,
linkcolor=porange,
urlcolor=pgreen
]{hyperref}

\usepackage{tikz}

\newcommand{\mean}[1]{\ensuremath{\langle #1 \rangle}}

\begin{document}

\title{Qubits in second quantisation in fermionic simulators}

\author{Ahana Ghoshal}
\thanks{These authors contributed equally to this work.}
\affiliation{Naturwissenschaftlich-Technische Fakult\"{a}t, Universit\"{a}t Siegen, Walter-Flex-Stra\ss e 3, 57068 Siegen, Germany}

\author{Carlos de Gois}
\thanks{These authors contributed equally to this work.}
\affiliation{Naturwissenschaftlich-Technische Fakult\"{a}t, Universit\"{a}t Siegen, Walter-Flex-Stra\ss e 3, 57068 Siegen, Germany}
\affiliation{CPHT, LIX, CNRS, Inria, École polytechnique, Institut Polytechnique de Paris, Palaiseau, France}

\author{Kiara Hansenne}
\affiliation{Naturwissenschaftlich-Technische Fakult\"{a}t, Universit\"{a}t Siegen, Walter-Flex-Stra\ss e 3, 57068 Siegen, Germany}
\affiliation{Universit\'{e} Paris-Saclay, CEA, CNRS, Institut de Physique Th\'{e}orique, 91191 Gif-sur-Yvette, France}

\author{Otfried G\"uhne}
\affiliation{Naturwissenschaftlich-Technische Fakult\"{a}t, Universit\"{a}t Siegen, Walter-Flex-Stra\ss e 3, 57068 Siegen, Germany}

\author{Hai-Chau Nguyen}
\affiliation{Naturwissenschaftlich-Technische Fakult\"{a}t, Universit\"{a}t Siegen, Walter-Flex-Stra\ss e 3, 57068 Siegen, Germany}

\begin{abstract}
    Simulating many-body fermionic systems in conventional qubit-based quantum computers poses significant challenges due to the overheads associated with the encoding of fermionic statistics in qubits, leading to the proposal of native fermionic simulators as an alternative. 
    While allowing for fermionic problems to be simulated efficiently, this class of fermionic simulators carries also specific constraints with them and poses other challenges unfamiliar to qubit systems. Here, we propose to pair fermionic modes to form a so-called  qubit in second quantisation representation. This allows fermionic gates to be represented as rotations of these second quantised qubits, enabling adaptation of methods for qubit systems.
    As an application, we use this pairing scheme to represent the measurement of two- and four-point correlators in fermionic simulators with its native gates as a graph problem.   
    Optimising measurement settings is then analysed with various analytical and algorithmic methods. 
\end{abstract}

\maketitle

{\it Introduction---}%
Quantum simulators~\cite{Georgescu2014,Altman2021,Paudel2022,Fraxanet2023} are designs replicating the behaviours of complex quantum systems, which can provide insights into phenomena that are difficult to study experimentally or computationally. 
Although universal digital quantum computers based on qubits \cite{Arute2019,Bluvstein2023,Kim2023,googleai2023} can, in principle, serve as quantum simulators, they face various challenges in practice. This is particularly the case for simulating fermionic systems, which are fundamental to quantum chemistry, condensed matter physics and materials science. 
Indeed, to simulate the fermionic exchange symmetry using qubits, complex encoding schemes such as the Jordan-Wigner~\cite{Jordan1928},
Bravyi-Kitaev mappings~\cite{Bravi2002}, or those based on a ternary tree structure~\cite{Jiang2020}, are required, introducing significant quantum gate overhead that is detrimental to current noisy qubit systems.  
 
To address these challenges, quantum simulators based on fermionic particles
with inherent fermionic statistics have been proposed.
Facilitated by recent breakthroughs in manipulating ultracold atoms in optical tweezer arrays~\cite{Norcia2018,Yan2022,Spar2022,González-Cuadra2023}, the development of such native fermionic quantum simulators has gained a significant momentum within the last few years~\cite{Argüello-Luengo2019,schuckert2024,tabares2025,gkritsis2025simulating,González-Cuadra2023}. In particular, fermionic gates formulated for neutral atom arrays demonstrate significant potential for advancing the implementation of fermionic quantum processors~\cite{González-Cuadra2023,Henriet2020,Evered2023,Wintersperger2023,Li2023}. 

While having natural fermionic exchange statistics, native fermionic simulators do have platform-specific constraints requiring attention. In particular, with the current generation of fermionic simulators, complex operations such as the coherent implementation of particle non-conserving transformations and Majorana transformations remain very challenging~\cite{Argüello-Luengo2019,González-Cuadra2023,gkritsis2025simulating}.
In general, the development of fermionic quantum simulators with all their platform-specific constraints raises various theoretical questions once faced by qubit-based quantum simulators. For example, can scalable techniques for state characterisation such as shadow tomography and overlapping tomography for qubit system be adapted to fermionic simulators? Or can error mitigation and error correction techniques be formulated directly for these simulators in a similar way to qubit systems?

\begin{figure} 
    \centering
    \includegraphics[width=.95\columnwidth]{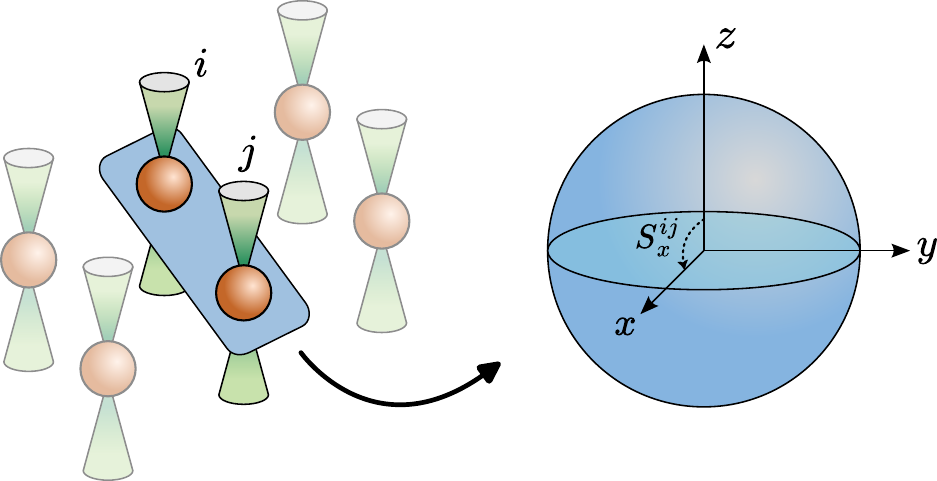}
    \caption{Schematic diagram of fermionic pairing used to define {the so-called qubit in second quantisation}. By grouping fermionic modes $i$ and $j$ into pairs, effective qubits are formed on which native fermionic gates act as qubit rotations. Here, $S_x^{ij}$ denotes the fermionic operator acting on the paired modes, corresponding to rotations on the Bloch sphere.}
    \label{fig:intro}
    \end{figure}

Here, we address part of this challenge by using the representation of qubit operations by fermionic modes. 
Specifically, we suggest pairing fermionic modes of the simulator to form so-called ``qubit in second quantisation,'' over which native gates act as qubit rotations; see Fig.~\ref{fig:intro}. 
This then allows for analytical methods to be developed in a very similar way as for qubit systems.

We demonstrate a direct application of this technique in 
the optimisation of the measurement of lower-order correlators, in particular two-point or four-point correlators (like $\mean{b_i^\dagger b_j}$ or $\mean{b_i^\dagger b_j b_k^\dagger b_l}$, respectively) in a quantum simulator.  
Since these correlators act over few modes, each of them can be measured by operating on the corresponding few modes. 
Different correlators may however overlap in their supporting modes and cannot be simultaneously measured in a single global measurement setting. 
The problem of fermionic overlapping tomography is then to choose these global measurements in an optimal way. This problem has been recently discussed in the literature using Majorana transformations~\cite{Jiang2020,Monroig2020,araujo2022local}. For contemporary fermionic simulators with only particle-number-conserving gates, it remains as an open challenge.

By pairing the modes, the fermionic correlators can be converted to correlators between {the qubits in second quantisation}, and the techniques of overlapping tomography for qubits~\cite{Cotler2020,Monroig2020,Maniscalco2020,verteletskyi2020measurement,araujo2022local,jaloveckas2023efficient, berg2024techniques,hansenne2024optimal,veltheim2024multiset} can then be applied. 
To this end, we develop a graphical framework representing not only the correlators between qubits alone such as used in Ref.~\cite{hansenne2024optimal}, but also the constraints imposed by the pairing scheme, turning the optimisation of measurement settings into a graph theoretic problem. 
Using graphical reasoning, we show that in a $n$-mode fermionic simulator, the necessary and sufficient number of measurement settings to determine all two-point correlators is linear in $n$ and provide an explicit construction.  
We then show that the minimum number of measurement settings to estimate all four-point correlators is proportional to $n^2$. This even reduces to constant numbers when the correlators are restricted to those defined on tiles of common lattices such as square, triangular, hexagonal, and Kagome ones.

{\it Fermionic simulators and their native gates---}%
A particularly promising approach to modelling fermionic quantum simulators with fermionic atoms involves the use of optical tweezer arrays. In the setup proposed by González-Cuadra et al.~\cite{González-Cuadra2023}, fermionic atoms are trapped individually in optical tweezers, forming a programmable array of fermionic modes. 
The platform supports two native gates. Tunnelling gates are described by 
 \begin{equation}
        \mathcal{U}_{ij}^{(t)}(\theta_1,\theta_2,\theta_3)=e^{-i\Big[{\theta_1}/{2}\big(e^{-i\theta_2}b_i^{\dagger}b_j+\text{H.c.}\big)+{\theta_3}/{2}\big(n_i-n_j\big)\Big]},
        \label{eq:elementary-gate}
    \end{equation}
and the interaction gates by
 \begin{equation}
        \mathcal{U}_{ij}^{(\text{int})}(\theta_4)=e^{-i\theta_4 n_i n_j}.
        \label{eq:elementary-gate-2}
    \end{equation}
Here, $b_i^\dagger$ and $b_i$ are fermionic creation and annihilation operators that satisfy $\set{b_i,b_j}=0$ and $\set{b_i^{\dagger},b_j}=\delta_{ij}$, with $n_i=b_i^{\dagger}b_i$ denoting the number operator. Notice that these elementary gates commute with the total number of fermions and form a universal gate set for particle-number-conserving unitaries~\cite{Bravi2002,González-Cuadra2023,gkritsis2025simulating,Li2023}.
Although operations altering the number of fermions or Majorana transformations cannot be coherently implemented, the fermionic simulator is expected to be effective for various problems in quantum chemistry and condensed matter physics~\cite{Argüello-Luengo2019,González-Cuadra2023,schuckert2024,gkritsis2025simulating,tabares2025}.

{\it Qubit representation in fermionic modes---}%
Basic to our approach is the possibility of representing qubit operators in fermionic modes~\cite{Altland_Simons_2010}. This allows qubit-based observables and protocols—typically defined in qubit systems—to be adapted to fermionic platforms without relying on mappings such as the Jordan-Wigner transformation.
To this end, for each Pauli matrix $\sigma_\alpha$ with $\alpha \in \set{x,y,z}$, one defines fermionic operators $S_{\alpha}^{ij}$ acting on fermionic modes $i$ and $j$ as
\begin{equation}
    S_{\alpha}^{ij} = 
    \begin{pmatrix}
        b_i^\dagger & b_j^\dagger
    \end{pmatrix}
    \frac{\sigma_\alpha}{2}
    \begin{pmatrix}
        b_i \\ b_j
    \end{pmatrix}.
\label{eq:spin-operators}
\end{equation}
One can easily confirm that the commutation relations of $S^{ij}_\alpha$ follow that from $\sigma_\alpha$. 
In fact, $S^{ij}_\alpha$ are known as spin operators in second quantisation representation and used frequently to investigate magnetic systems~\cite{Altland_Simons_2010}. 
By analogy, we will say that the pair of modes $(i,j)$ forms a qubit in second quantisation. 

In this language, the tunnelling gates~\eqref{eq:elementary-gate} represent general rotations of second quantised qubits,
\begin{equation}
 \mathcal{U}_{ij}^{(t)} (\theta_1,\theta_2,\theta_3) = \exp\{-i (\phi_x S^{ij}_{x}+ \phi_y S^{ij}_{y}+ \phi_z S^{ij}_{z})\},   
\end{equation}
where $\phi_x= \theta_1 \cos \theta_2 $, $\phi_y= \theta_1 \sin \theta_2 $ and $\phi_z= \theta_3 $, and the interaction gate reads
\begin{equation}
 \mathcal{U}_{ij}^{\rm (int)} (\theta_4) = \exp\{-i \gamma [(S_0^{ij})^2 -(S_z^{ij})^2]\} , 
\end{equation}
where $S_0^{ij}= (b_i^\dag b_i + b_j^\dag b_j)/2$, 
$S_z^{ij} = (b_i^\dag b_i - b_j^\dag b_j)/2$ and $\gamma=\theta_4$.

Notice that $S_0^{ij}$ and $S_z^{ij}$ can be computed directly from measurement of occupation numbers. On the other hand, $S_x^{ij}$ and $S_y^{ij}$ are related to the two-point correlators by $S_x^{ij} = (b_i^\dagger b_j + b_j^\dagger b_i)/2$ and $S_y^{ij} = i(b_j^\dagger b_i - b_i^\dagger b_j)/2$.     
To measure $S_x^{ij}$ and $S_y^{ij}$, we make use of the usual ``qubit rotations'' that relate them to $S_z^{ij}$, $S_x^{ij} = -e^{-i \theta S_y^{ij}} S_z^{ij} e^{+i \theta S_y^{ij}}$ and $S_y^{ij}= e^{-i \theta S_x^{ij}} S_z^{ij} e^{+i \theta S_x^{ij}}$, where $\theta = \pi / 2$; see also Appendix~\ref{app:rotation}.  
Therefore, all such second quantised Pauli observables can be eventually obtained by combined use of tunnelling gate and occupation number measurement, and two point correlation functions can be computed as  $\langle b_i^\dag b_j \rangle = ({\langle{S_x^{ij}} \rangle + i \langle{S_y^{ij}} \rangle})$.

{\it Measuring two-point correlators: graph representation---}%
Our aim is to estimate two-point correlation functions such as $\mean{b_i^\dagger b_j}$ for certain pairs of modes $(i,j)$, using only a single layer of native operations~\eqref{eq:elementary-gate} and~\eqref{eq:elementary-gate-2} from the fermionic simulator, and the minimal number of measurement settings.
As mentioned, with such a gate set, implementation of protocols based on number non-conserving operations as well as Majorana transformations is not straightforward, and we are to use the second quantised qubit mapping~\eqref{eq:spin-operators} outlined above. Note that for $i = j$, ${b_i^\dagger b_i}$ correspond to number operators $n_i$, which can be simultaneously measured with a single setting. In the following, we therefore assume $i \ne j$.

To determine the measurement settings, we first represent the correlation functions by a graph $G$ in which the vertices $i \in \set{ 1, \ldots, n}$ correspond to the modes of the fermionic system, and an undirected edge $(i, j)$ represents a term $b_i^\dagger b_j$ of interest and also its conjugate $b_j^\dagger b_i$. 
From the previous discussion, we know that any edge $(i, j)$ in $G$ can be determined from two measurement settings, ${S}_x^{ij}$ and ${S}_y^{ij}$.  
However, the measurements of ${S}_\alpha^{ij}$ and ${S}_\beta^{kl}$ ($\alpha, \beta \in \set{x, y}$) can only be performed in parallel when the corresponding edges do not share a vertex, as the necessary rotations would otherwise act on overlapping modes. 
A straightforward solution is to consider a colouring of the edges of $G$ such that no two edges that are incident to the same vertex share a colour. 
Each colour then denotes a set of independent pairs of modes, over which the measurements can be parallelised.
For each colour, we apply the measurement settings $S_x^{ij}$, and then $S_y^{ij}$, to all pairs of modes $(i,j)$ in the corresponding set, thereby requiring two distinct settings per colour. Letting $\chi^\prime(G)$ denote the minimum number of edge colours, it follows that obtaining all two-point correlators (with $i \ne j$) requires at most $2\chi^\prime(G)$ measurement settings, plus one additional setting to measure the occupation number operators.

    Edge colouring is well understood in graph theory. Vizing-Gupta's theorem \cite{Vizing1964, gupta1968studies} states that, for any graph, $\Delta(G) \leq \chi^\prime(G) \leq \Delta(G) + 1$, where $\Delta(G)$ is the maximum number of edges incident to a vertex in the graph, also known as its degree%
    ~\footnote{In general, the problem of determining whether $\chi^\prime$ is $\Delta$ or $\Delta + 1$ is NP-complete \cite{Holyer1981,Garey1983}. On the other hand, as the optimality gap is at most one, not much is lost by taking the upper bound, which would incur at most two additional measurement settings. Crucially, in that case it is also easy to actually construct a colouring of this degree, and therefore to find which measurements to perform. This can be done for any graph in time complexity $O(\abs{V} \abs{E})$ via the Misra-Gries construction \cite{MISRAGRIES}, which can be improved for particular classes of graphs (see e.g.\ \cite[Table 1]{Cole2008} for an overview of edge colouring algorithms for planar graphs).}.
    For many physically meaningful cases, $\chi^\prime (G)$ is independent of the number of modes in the system. For example, $\Delta = 4$ for a square lattice, therefore only $9$ measurement settings are sufficient to fully characterise any nearest-neighbour fermionic observable. 
    Most importantly, the colouring can usually be explicitly constructed and always gives a sufficient measurement scheduling. 
    As a concrete example, in Appendix~\ref{app:colouring-kn} we analyse the case where all two point correlation functions on $n$ fermionic modes are to be estimated, and find that this requires $2n-1$ or $2n + 1$ measurement settings for even or odd $n$, respectively.

Notice that there is a rich literature on overlapping tomography of fermionic systems using the Majorana transformation~\cite{Monroig2020,araujo2022local,Jiang2020}. In particular, it was demonstrated~\cite{Monroig2020} that $2n-1$ measurement circuits {suffice to reconstruct all two-point correlators.}  More generally, a graph-colouring technique has been used to show that $2\chi^{\prime}(G)+1$ 
measurement settings are required for the reconstruction of the two-point correlators associated to $G$ \cite{araujo2022local}.
Notably, our scheme allows for contemporary fermionic simulators without Majorana transformations to also achieve the same scaling in the number of measurement settings.

{\it Measuring four-point correlators---}%
Although two-point correlation functions can capture key properties of fermionic systems, the interaction between particles often involves higher-order correlation functions. 
Incorporating four-point correlations of the form $\mean{b_i^\dagger b_j b_k^\dagger b_l}$ into the measurement procedure would enable the description of other properties of interest in chemistry and condensed matter physics, including ground-state properties and excitation energies \cite{loewdin1955quantum,Coleman1963}. 
We are again to see that the techniques of qubits in second quantisation allow us to address this problem for contemporary fermionic simulators graphically.

Due to the Pauli exclusion principle, the four-point correlators $\mean{b_i^\dagger b_j b_k^\dagger b_l}$ for which $i = k$ or $j = l$ vanish. 
We classify the remaining four-point correlators on modes $i$, $j$, $k$, $l$ into three types: (i) two repeated indices ($i=j$ and $k=l$), resulting in $\mean{b_i^\dagger b_i b_j^\dagger b_j} = \mean{n_in_j}$; (ii) one repeated index ($i=j$), resulting in $\mean{b_i^\dagger b_i b_j^\dagger b_k} = \mean{n_i b_j^\dagger b_k}$; and (iii) all indices are distinct, leading to $\mean{b_i^\dagger b_j b_k^\dagger b_l}$.
Clearly, to obtain the correlators of type (i), it suffices to measure the number operators $n_i$ and $n_j$ simultaneously. 
For correlators of type (ii), measuring $n_i S_x^{jk}$ and $n_i S_y^{jk}$ is sufficient to recover their values, which follows from the basic measurement procedure for two modes discussed previously. 
Lastly, correlators of type (iii) can be written as $4 b_i^\dagger b_j b_k^\dagger b_l = S_x^{ij} S_x^{kl} - S_y^{ij} S_y^{kl} + i \left( S_x^{ij} S_y^{kl} + S_y^{ij} S_x^{kl} \right)$, thus can be obtained by measuring the second-quantised qubit correlations $S_x^{ij} S_x^{kl}$, $S_x^{ij} S_y^{kl}$, $S_y^{ij} S_x^{kl}$, and $S_y^{ij} S_y^{kl}$.

    Given any set of four-point correlators that we want to determine, our goal is then to find a small set of measurements composed of products of $n_i$, $S_x^{jk}$, and $S_y^{lm}$ that give access to the desired correlators.
    To do so, we will frame the solution as a certain class of graph covering problems called edge clique covering \cite{orlin1977contentment,kou1978covering}. This combinatorial structure is central in software testing \cite{Hartman2005,torres2013survey} and has previously appeared in the context of overlapping tomography of qubits \cite{jaloveckas2023efficient, berg2024techniques,hansenne2024optimal}.

\begin{figure} 
    \centering
        \subfigure[]{\includegraphics[width=.45\columnwidth]{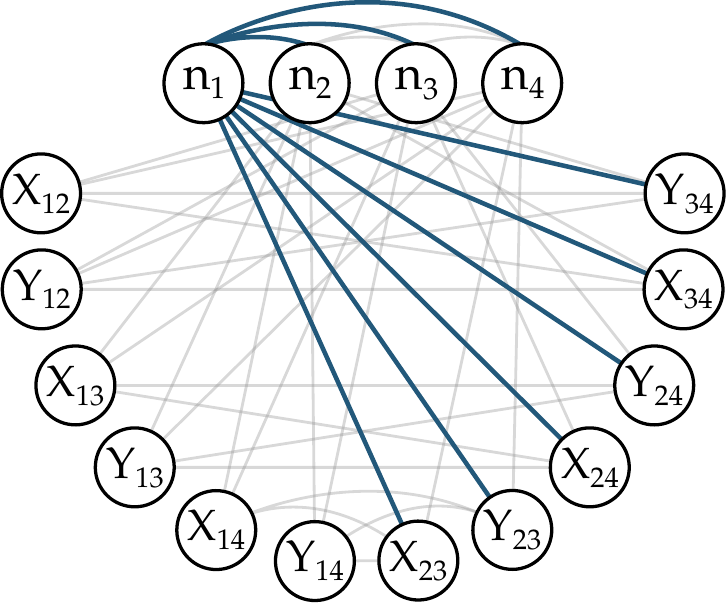}\label{fig:gm-full}}\hspace{1.5em}
        \subfigure[]{\includegraphics[width=.45\columnwidth]{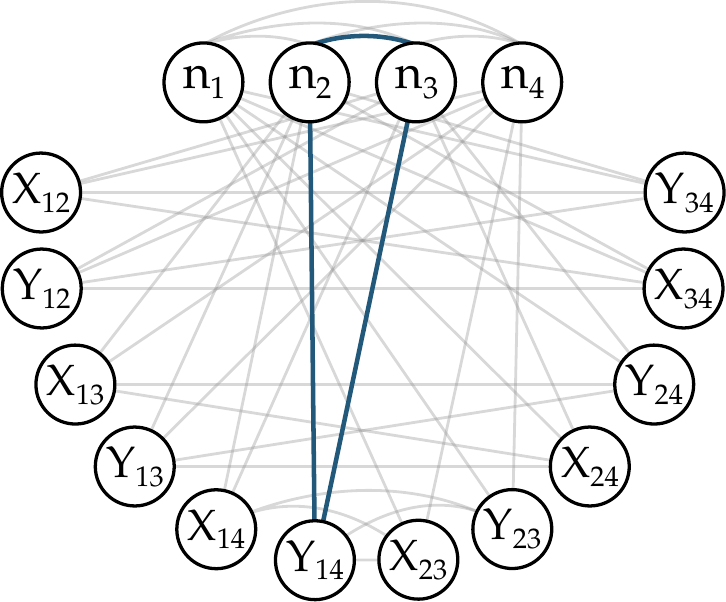}\label{fig:gm-clique}}
    \caption{The measurement graph $G_M$ for a four-mode system, where as a shorthand we represent the operators $S_{x}^{ij}$ and $S_{y}^{ij}$ as $X_{ij}$ and $Y_{ij}$, respectively.
    Each node represents a one- or two-mode operator.
    Edges connect nodes if the corresponding operators can be measured in a single setting, i.e., if they do not share indices.
    Highlighted in thick, blue lines in (a) are all the edges incident to $n_1$, and in 
    (b) a clique of $G_M$, corresponding to the measurement $n_2n_3S_y^{14}$. }
    \label{fig:gm}
    \end{figure}

{\it Graph representation of fermionic mode pairing constraints and four-point correlators---}%
Assuming again one layer of elementary gates, we start by constructing a graph $G_M$ to represent the feasible measurements in the fermionic simulator. 
    In this graph, a vertex corresponds to either a number operator $n_i$ or one of the operators $S_x^{ij}$ or $S_y^{ij}$ (with $i\neq j$). Since
    $S_x^{ji} = S_x^{ij}$ and $\quad S_y^{ji} = - S_y^{ij}$, it is sufficient to only consider the pairs of modes $(i,j)$ with $i < j$.
    Any two vertices in $G_M$ will be connected if the corresponding measurements can be performed in parallel. This means that we draw an edge between all vertices $n_i$, but $n_i$ and $S_\alpha^{jk}$ will be connected if and only if $i \neq j$ and $i \neq k$. Lastly, the vertices representing $S_\alpha^{ij}$ and $S_\alpha^{kl}$ are only connected when all four indices $i,j,k,l$ are distinct; see Fig.~\ref{fig:gm-full}.
    Fully connected subgraphs of $G_M$ (called \textit{cliques}) correspond to measurements that can be performed in parallel; see Fig.~\ref{fig:gm-clique}.
    
    We then construct a target graph $G_T$ that represents the four-point correlators under interest.
    The set of vertices of $G_T$ is the same of $G_M$. 
    For each four-point correlator that we want to obtain, we connect the vertices of $G_T$ associated with the measurements that must be performed to do so. 
    For example, correlators of the form $\mean{n_in_j}$ require us to connect the vertices $n_i$ and $n_j$ in $G_T$. 
    For correlators of the form $\mean{n_i b_j^\dagger b_k}$, two edges are necessary: one between $n_i$ and $S_x^{jk}$, and another between $n_i$ and $S_y^{jk}$. 
    Due to the labelling of the vertices for $j > k$, 
    the corresponding vertices are $S_x^{kj}$ and $S_y^{kj}$. 
    Therefore, both $\mean{n_i b_j^\dagger b_k}$ and $\mean{n_i b_k^\dagger b_j}$ are represented by the same two edges. This is justified by the fact that both correlators can be obtained from the same measurement settings, as they are complex conjugates of each other. 
    Lastly, for four-point correlators with distinct indices we draw four edges, corresponding to the four measurements $S_\alpha^{ij} S_\beta^{kl}$ for $\alpha, \beta \in \set{x, y}$ necessary to reconstruct the correlator.
    
    With $G_M$ and $G_T$ in hand, finding a small set of measurements that allow for the reconstruction of the target correlators then becomes the following graph problem: Find a small set of cliques of $G_M$ (the possible measurements) that cover every edge of $G_T$ (the required measurements to determine every observable of interest). 
    This is a variant of the clique covering problem in graph theory \cite{orlin1977contentment,kou1978covering}, which has already proven useful for optimising measurement settings for quantum information tasks \cite{verteletskyi2020measurement, jaloveckas2023efficient, berg2024techniques, veltheim2024multiset, hansenne2024optimal}.
    For small instances, the minimum solution can be found exactly, for example using backtracking algorithms or binary programming \cite{hansenne2024optimal}. Otherwise, one can turn to a variety of heuristic procedures that are efficient and can in many cases find good solutions (see, e.g.~\cite{gramm2006datareduction,conte_clique_2016} and references therein). In Appendix~\ref{app:covering-algorithms} we detail
    formulations of an exact and a heuristic algorithm for this problem.

{\it Application I: All four-point correlators---}%
To start, notice that the graph $G_M$ is simply determined by the number of modes in the system and is thus independent of which correlators we wish to learn. 
It is thus just left to determine the graph $G_T$ corresponding to the case of all four-point correlators.
Notably, due to the canonical anticommutation relations of fermionic creation and annihilation operators, many four-point correlators are equal, sometimes up to a sign.
Let us go again through the different cases.
When two indices are repeated, it is clear that $\mean{n_in_j} = \mean{n_j n_i}$, so without loss of generality we consider that $i < j$.
    When no index is repeated, we notice that $\mean{b_i^\dagger b_j b_k^\dagger b_l} = - \mean{b_k^\dagger b_j b_i^\dagger b_l} = - \mean{b_i^\dagger b_l b_k^\dagger b_j}=\langle b_k^{\dagger}b_l b_i^{\dagger} b_j\rangle$, so we only need to recover $\mean{b_i^\dagger b_j b_k^\dagger b_l}$ for $i < k$ and $j < l$.
    Incorporating the symmetries of $S_x^{ij}$ and $S_y^{ij}$, we conclude that to recover all the four-point correlators of an $n$-mode fermionic system, it is sufficient to consider the four-point correlators (i) $\mean{n_in_j}$ with $i < j$; (ii) $\mean{n_i b_j^\dagger b_k}$ with $j < k$; and (iii) $\mean{b_i^\dagger b_j b_k^\dagger b_l}$ with $i < k < l$ and $i < j < l$, where $i,j,k,l \in \set{1,2,\ldots, n}$ are distinct.

\begin{figure} 
    \centering
    \includegraphics[width=.95\columnwidth]{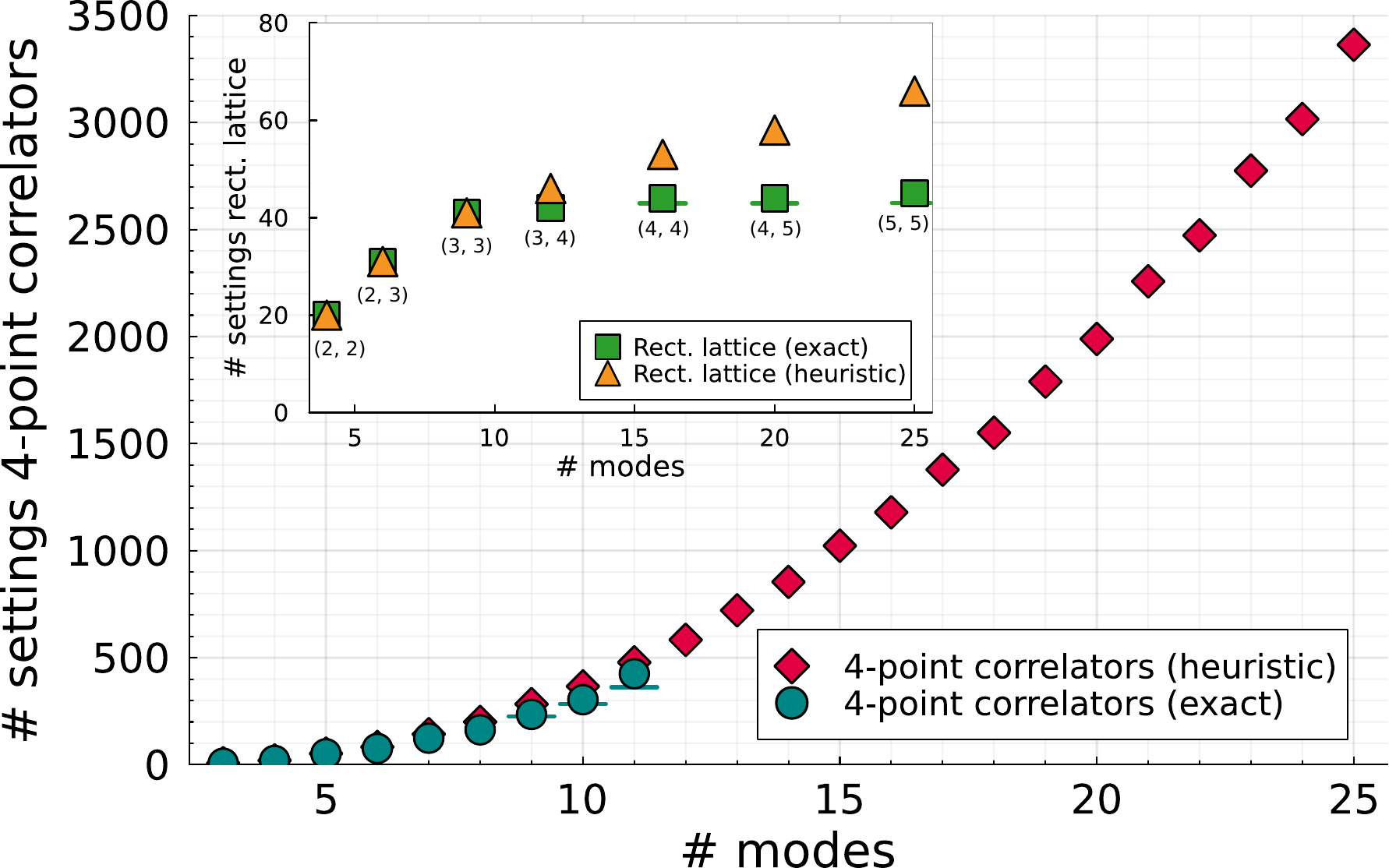}
    \caption{Number of settings for learning every four-point correlator (main figure) and nearest-neighbour correlators on rectangular lattices of size $(x, x)$ and $(x, x + 1)$ (inset), as a function of the number of modes $n$. For the inset, $n=x^2$ for $(x,x)$ and $n=x(x+1)$ for $(x,x+1)$.
    In the main figure, while the exact minimum could only be computed for up to eight modes, the heuristic algorithm provides works for much larger system sizes.
    The horizontal bars indicate cases where convergence was not guaranteed, and a smaller solution down to the bar may exist.}
    \label{fig:nof-settings-covering}
    \end{figure}

    After translating independent four-point correlators into the target graph $G_T$, we obtained the solutions through an exact and a heuristic edge clique cover method, for different number of modes (Appendix~\ref{app:covering-algorithms}).
    While the exact method builds a minimal solution via binary programming but is not scalable, the heuristic provides suboptimal solutions but scales efficiently. 
    The resulting number of measurement settings are shown in Fig.~\ref{fig:nof-settings-covering} (see also Appendix~\ref{app:example-settings-four-point-kn} 
    for explicit solutions to selected cases).

    Notice that a simple lower bound on the number of settings follows from the fact that the $4(n-1)(n-2)$ elements from $\set{ S_x^{1i} S_x^{2 j}, S_x^{1i} S_y^{2 j}, S_x^{1i} S_y^{2 j}, S_x^{1i} S_y^{2 j}  }$ with $j \in \set{1, 3, \dots, n} \setminus \set{i}$ and $i \in \set{2, \dots, n}$ must be measured, however none can be measured in parallel.
    Therefore, the minimum number of settings grows at least as fast as $n^2$. 
    Using Majorana transformations, a protocol with a scaling proportional to $n^2$ has been achieved~\cite{Monroig2020}.

 \begin{figure}[t] 
    \centering
        \subfigure[]{\includegraphics[width=.43\columnwidth]{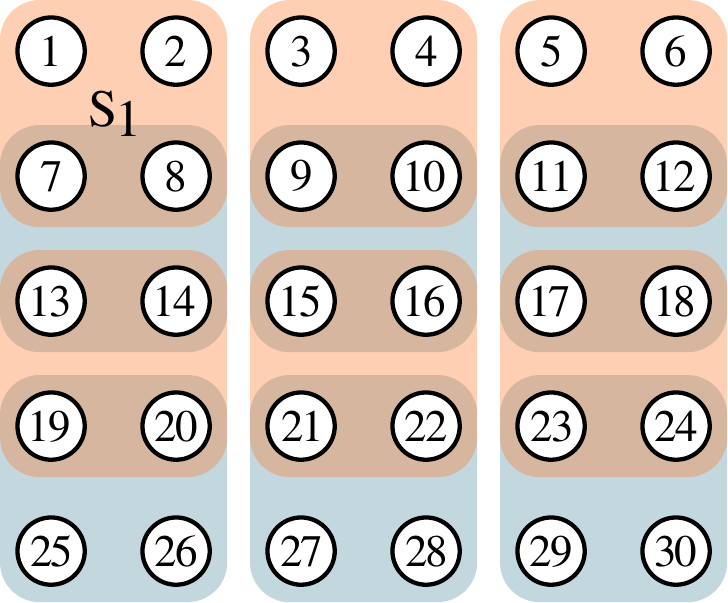}\label{fig:tiling-1}}\hspace{1.2em}
        \subfigure[]{\includegraphics[width=.43\columnwidth]{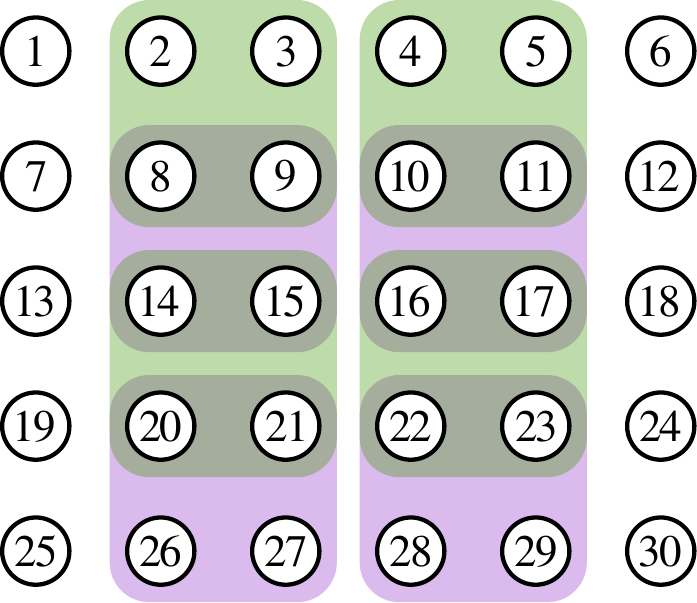}\label{fig:tiling-2}}
    \caption{Non-overlapping tiles of a $5 \times 6$ rectangular lattice [the same lattice is depicted in \subref{fig:tiling-1} and \subref{fig:tiling-2}]. The four-point correlators involving modes in each shaded region can be determined with the $20$ measurement settings for the four-mode correlators (Appendix \ref{app:example-settings-four-point-kn}). Modes in tiles of the same colour can be measured in parallel, using the same settings on the respective modes. Since any rectangular lattice can be covered with four different colours, all nearest-neighbour four-point correlators can be determined with $80$ settings.}
    \label{fig:square-4pc}
    \end{figure}   

{\it Application II: Planar tilings---}
    In lattice-based fermionic systems with local interactions, it is often sufficient to perform tomography on small regions, such as the four-mode tiles of a square lattice. By focusing on four-point correlators within each tile, one captures essential local properties while simplifying the measurement process. For a wide class of geometries, including square, triangular, hexagonal, and Kagome lattices, the four-point correlators can be extracted using a constant number of measurement settings, independent of the system size.
    
    The reason for that is best explained by an example. Consider the $5 \times 6$ square lattice depicted in Fig.~\ref{fig:square-4pc}.
    From our previous discussion, we know how to determine every four-point correlator containing modes $1$, $2$, $7$, $8$ (region $S_1$ in Fig.~\ref{fig:square-4pc}) with $20$ measurement settings (cf.\ Appendix~\ref{app:2rdm-k4-settings}).
    Since the modes contained in the other regions with the same colour in Fig.~\ref{fig:square-4pc} do not overlap with the ones in $S_1$, they can be measured in parallel, thus without increasing the number of settings. This is, however, not true for the regions depicted with different colours, which can themselves be measured in parallel, but not together with the previous ones. It is easy to convince oneself that this partitioning of the modes into four colours extends to any rectangular lattice, therefore a constant number of $80$ measurement settings is sufficient, independently of the system size.
    Although constant, this turns out to be not optimal. In fact, as can be seen in Fig.~\ref{fig:nof-settings-covering}, the exact numerical solutions suggest a significantly smaller plateau at around $45$ settings.

\begin{table}[t!!]
        \centering
        \begin{tabular}{ccccc}
            
          Lattice&  \,square\, & \,triangular\, & \,hexagonal\, & \,kagome\, \\ \hline
            \hline 
          \, \# settings \,   & $80$ & $42$ & $228$ & $166$
        \end{tabular}
        \caption{ 
        Constant number of settings for obtaining all four-point correlators of the modes involved in each tile of the lattice.}
        \label{tab:lattices}
    \end{table}

    The same reasoning can be used to show that other common lattices can be measured in a constant number of settings.
    For example, the basic measurement for a triangle can be done with $7$ settings (see Appendix~\ref{app:2rdm-k3-settings}), 
    and six colours are needed to cover the lattice with non-overlapping triangles, thus $42$ settings are sufficient in this case.
    Similarly, the hexagonal lattice is made of six-mode tiles that can be grouped in three sets of non-adjacent tiles. As obtaining all four-point correlators of six modes requires $76$ settings, $228$ global settings are sufficient.
    Lastly, the Kagome lattice is made of triangular and hexagonal tiles, and obtaining all four-point correlators of each three- and six-mode tiles can be done with $166$ global settings.
    Table \ref{tab:lattices} summarizes these examples. The estimation of correlators between neighbouring modes has also been studied for fermionic systems with Majorana transformations \cite{araujo2022local}, where a similar counting procedure was used.

{\it Conclusion.---}%
    By pairing fermionic modes to form {qubits in second quantisation}, we paved the way for mapping methods for qubit-based quantum processors to that for the fermionic ones in a straightforward manner. This was illustrated by obtaining protocols to measure fermionic correlators using methods of qubit overlapping tomography. Our work opens a range of questions for further investigation. While we have considered the fermionic simulator, it is also natural to consider the applications of second-quantised qubits in bosonic systems. Investigation of the mapping of different quantum information protocols and algorithms for qubits other than the overlapping tomography is naturally a promising direction. As using qudits are sometimes advantageous over qubits~\cite{Wang_2020}, using representations of higher spins than the qubits could be expected to bring reduction in the complexity of various protocols.

{\it Acknowledgments---}%
We thank Malte Lochau and Mathis Wei{\ss} 
 for discussions.
This work was supported by the Deutsche Forschungsgemeinschaft 
(DFG, German Research Foundation, project number 563437167), the 
Sino-German Center for Research Promotion (Project M-0294), the 
German
Federal Ministry of Research, Technology and Space (Project QuKuK, Grant No. 16KIS1618K and Project
BeRyQC, Grant No. 13N17292), the House of Young Talents of the University of Siegen, the project EINQuantum NRW, and the European Union's Horizon 2020 Research and Innovation Programme under QuantERA Grant Agreement no.\ 731473 and 101017733. A. Ghoshal acknowledges the support from the Alexander von Humboldt Foundation. 


\begin{appendix}

\section{Measurement of $S_x^{ij}$ and $S_y^{ij}$}
\label{app:rotation}

To measure $S_x^{ij}$ and $S_y^{ij}$, we make use of the usual ``qubit rotations'' that relate them to $S_z^{ij}$. Specifically, we use the following rotation relations
    \begin{eqnarray}
        &&e^{i\theta S_y^{ij}}S_z^{ij} e^{-i\theta S_y^{ij}}=S_z^{ij}\cos\theta-S_x^{ij}\sin\theta=-S_x^{ij},\nonumber\\
&&e^{i\theta S_x^{ij}}S_z^{ij} e^{-i\theta S_x^{ij}}=S_z^{ij}\cos\theta+S_y^{ij}\sin\theta=S_y^{ij},\;\;\;\;
\label{eq:rotaion}
    \end{eqnarray}
    with $\theta=\pi/2$. These identities hold as consequences of the commutation relations of the Pauli operators as generators of the rotational group. To be precise, by expanding the left-hand sides of both the equations of Eq.~(\ref{eq:rotaion}) by the Baker–Campbell–Hausdorff (BCH) expansion, we get
    \begin{eqnarray}
        e^{i\theta S_y^{ij}}S_z^{ij} e^{-i\theta S_y^{ij}}= \sum_{m=0}^{\infty}\frac{(i\theta)^m}{m!}[S_y^{ij},S_z^{ij}]_m,
        \label{eq:BCH}
    \end{eqnarray}
    where $[S_y^{ij},S_z^{ij}]_m=\underbrace{[S_y^{ij},\cdots[S_y^{ij},[S_y^{ij}}_{m\; \text{times}},S_z^{ij}]]$ and $[S_y^{ij},S_z^{ij}]_0=S_z^{ij}$.
    Hence, Eq.~(\ref{eq:BCH}) becomes
    \begin{equation}
        e^{i\theta S_y^{ij}}S_z^{ij} e^{-i\theta S_y^{ij}}= S_z^{ij}+i\theta [S_y^{ij},S_z^{ij}] - \frac{\theta^2}{2!} [S_y^{ij},[S_y^{ij},S_z^{ij}]]+\cdots
    \end{equation}
Using the standard spin commutation relations
\begin{equation}
[S_\alpha^{ij},S_\beta^{ij}]=i\epsilon_{\alpha \beta \delta}S_\delta^{ij},
\end{equation}
where $\epsilon_{\alpha \beta \delta}$ is the Levi-Civita symbol, defined by its cyclic property over $(x,y,z)$ as
$$\epsilon_{xyz}=\epsilon_{yzx}=\epsilon_{zxy}=1,\quad\epsilon_{yxz}=\epsilon_{xzy}=\epsilon_{zyx}=-1,$$ 
we obtain
\begin{equation}
    e^{i\theta S_y^{ij}}S_z^{ij} e^{-i\theta S_y^{ij}}=S_z^{ij}\cos\theta-S_x^{ij}\sin\theta.
\end{equation}
A similar calculation yields
\begin{equation}
    e^{i\theta S_x^{ij}}S_z^{ij} e^{-i\theta S_x^{ij}}= S_z^{ij}\cos\theta+S_y^{ij}\sin\theta.
\end{equation}
Now, by choosing $\theta=\pi/2$, we recover both the identities of Eq.~(\ref{eq:rotaion}), which are used to rotate the spin measurement basis from $S_z^{ij}$ to $S_x^{ij}$ and $S_y^{ij}$, respectively.

\section{Measurement settings for all two-point correlators }
    \label{app:single-body-settings}
    \label{app:colouring-kn}

 As a concrete example, we examine the estimation of all two-point correlation functions among $n$ fermionic modes.
        Any two-point correlator can be estimated if $\langle b_i^\dagger b_j \rangle$ is known for all pairs or modes $i,j$.
        From our previous discussion, this requirement can be represented by the complete graph $G = K_n$, and a colouring of its edges gives us a way to parallelise the measurements. 
       The optimal colouring of complete graphs is completely known in graph theory, and is described here for completeness.

\begin{figure}
    \centering
        \subfigure[]{\includegraphics[width=.4\columnwidth]{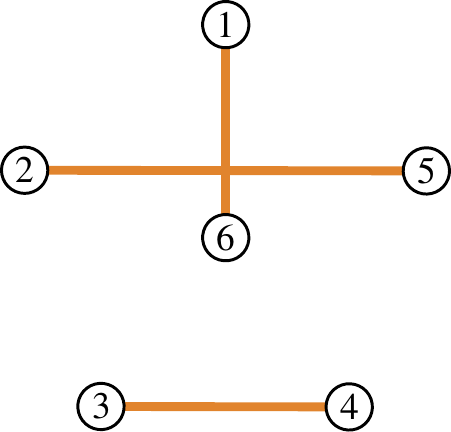}\label{fig:colouring-k6-single}}\hspace{1.5em}
        \subfigure[]{\includegraphics[width=.4\columnwidth]{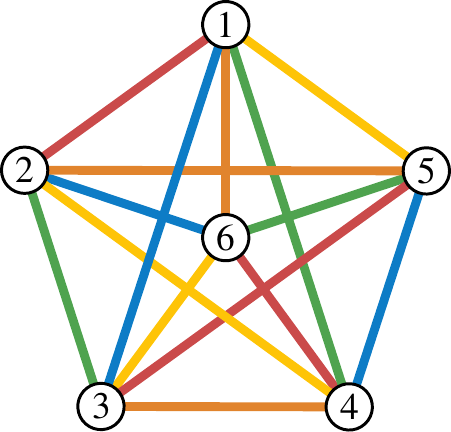}
        \label{fig:colouring-k6-rotated}}
    \caption{Colouring of the $K_6$ graph using $5$ colours. 
    In \subref{fig:colouring-k6-single} we pick a colour for the radial edge $\set{1, 6}$ then draw all edges that are perpendicular to it. 
    The second step \subref{fig:colouring-k6-rotated} is to rotate this construction using a different colour. 
    This solution is optimal and the procedure can be naturally extended to any complete graph with an even number of vertices.
    Each colour corresponds to two settings; for \subref{fig:colouring-k6-single}, these settings are $S_x^{16} S_x^{25} S_x^{34}$ and $S_y^{16} S_y^{25} S_y^{34}$. 
    \label{fig:colouring-k6}
    }
    \end{figure}

        Let us first consider the case of $K_{2n}$ and a set of colours $C = \set{ c_k }_{k=1}^{2n-1}$.
        The construction can be visualised geometrically in the complex plane. 
        We start by distributing the vertices $v \in \set{1, \ldots, 2n - 1}$ around the unit circle, at the positions $e^{v \frac{2 \pi i}{2n-1}}$. 
        The remaining vertex ($v = 2n$) goes at the origin.
        Edges connecting the central vertex to any other are of the form $\set{ v, 2n }$, for $v$ in $\set{1, \ldots, 2n - 1}$, and they must be given different colours since they meet at $2n$. This can be satisfied by attributing colour $c_v$ for each edge of this form. At this point, only $2n - 1$ edges are coloured. To continue, fix any $v \in \set{1, \ldots, 2n - 1}$ and consider the edges of the form 
        $\set{v+\ell \mod (2n-1),f(v,\ell)}$ where
        \begin{eqnarray}
        f(v,l)=
        \begin{cases}
        (v-l) \mod (2n-1) \quad \text{if}\; (v-l)\ne 0 \\
         2n-1 \quad \quad \quad \quad \quad \quad \quad \quad \text{if}\; (v-l)=0
        \end{cases}
        \end{eqnarray} 
       and $\ell \in \set{ 1, \ldots, n - 1 }$. 
       From their angles, it is easy to see these edges are perpendicular to $\set{ v, 2n }$. Therefore they are parallel and never meet at any vertex, thus it is safe to colour them with $c_v$ (see Fig.~\ref{fig:colouring-k6-single} for an illustration of this step).

        Proceeding analogously for the other choices of $v$ (Fig.~\ref{fig:colouring-k6-rotated}) and noticing that for any $v^\prime \neq v$, the edges 
        $\set{v+\ell \mod (2n-1),f(v,\ell)}$ and $\set{v^{\prime}+\ell \mod (2n-1),f(v^{\prime},\ell)}$ are distinct, there are now $(n - 1) (2n - 1)$ newly coloured edges. Summing up with the initial ones, we get $n (2n - 1)$ distinct edges. This is equal to the total number of edges in $K_{2n}$, and the colouring is valid by construction. For the case of an odd number of vertices ($G = K_{2n - 1}$), we can simply colour the graph $K_{2n}$ and then remove the extra vertex and its associated edges. We are left with $2n-1$ colours, which is nevertheless optimal \cite{handbook}.
        
        Similar constructions appear in many forms across the combinatorial designs literature, where they are also presented as the $1$-factorisability of complete graphs, as the ``round-robin tournament'' problem or as symmetric Latin squares; see for example \cite[Chap.\ 5.5]{handbook} for a more elegant construction using group theoretic methods. As can be expected, this is not the only possible edge colouring $K_n$. In fact, the number of non-isomorphic colourings for $n = 4, 6, 8$ and $10$ are known to be $1, 6, 6240$ and $1225566720$ \cite{oeisA000438}. 
        
        Most importantly to us, it is now straightforward to construct a sufficient and economical measurement scheduling for learning the two-point correlators of any fermionic system with $n$ modes:
        
    \begin{algorithm}
    \caption{Measuring $\set{ \langle b_i^\dagger b_j \rangle}_{i,j=1}^n$ (with $n$ even)}
    \label{proc:single-body-kn}
    \SetAlgoLined
    \nl measure $n_1 \ldots n_n$\; 
    \For{each $v \in \set{1, \ldots, n-1}$}{
        \nl select modes 
        $m_v = \set{\set{v, n}} \cup \set{\set{v + \ell \mod (n-1), f(v,\ell)}}_{\ell = 1}^{n/2-1}$ \;
        \nl apply $S_x$ to each pair in $m_v$ and measure $n_1 \ldots n_n$\;
        \nl apply $S_y$ to each pair in $m_v$ and measure $n_1 \ldots n_n$\;
    }
    \end{algorithm}
      

\section{Methods to solve graph covering problems}
\label{app:covering-algorithms}
    The variants of the edge clique covering problem mentioned in the main text can usually only be solved to optimality in small instances, while for larger instances one usually resorts to heuristic methods \cite{orlin1977contentment,kou1978covering,gramm2006datareduction,conte_clique_2016}. Here we describe the two exact and one heuristic solutions that were used in this work (cf.\ Fig.~\ref{fig:nof-settings-covering} of the main text). 
    Notice that this problem is closely related to that of combinatorial software testing, for which a wide array of different approaches also exist \cite{torres2013survey,Hartman2005}.
    
    For the first exact formulation, suppose a graph $K$ and its set of maximal cliques $C_K$ are given. With a minimal amount of these cliques, we want to cover the edges {{$E(G)$}} of another given graph $G$. 
    {Note that this problem is only feasible when the edge set of $G$ is a subset of the one of $K$.}
    Let $z_c$ be binary variables, where $c$ is a label for each clique in $C_K$. The value of $z_c$ represents whether clique $c$ is part of the solution ($z_c = 1$) or not ($z_c = 0$).
    For each edge $e$ in $G$ and each clique $c$, let us define $e_c = 1$ is $e$ belongs to $c$, otherwise $e_c = 0$.
    When $z_c = 1$, every edge of $G$ that is contained in clique $c$ can be considered covered, and this is represented by $z_c e_c = 1$.
    On the other hand, we also want that every edge of $G$ is in at least one of the cliques in the solution, and for that we can require that $\sum_c z_c e_c \geq 1$.
    Altogether, these can be written as:
    \begin{subequations}\label{app-eq:mip-given-cliques}
    \begin{align}
        \text{minimize: }   &\;\;\;\sum_c z_c \\
        \text{subject to: } &\;\;\;z_c \in \set{0, 1}, \quad \forall c \in C_K \\
                            &\;\;\;\sum_c z_c e_c \geq 1, \quad \forall e \in E(G)
    \end{align}
    \end{subequations}
    Notice that the values of $e_c$ are constants defined by the graphs $K$ and $G$. Therefore the program above is simply a binary linear program, which can be solved to global optimality.
    
    The model in Eq.~\eqref{app-eq:mip-given-cliques} was used to find the exact solutions for the problem of obtaining every four-point correlator. 
    For the rectangular lattices, though, the fact that there are exponentially many cliques in $K$ (and thus variables $z_c$), but very few are necessary to cover the corresponding $G$, suggests this might not be the most practical formulation. 
    In these cases, we have used a variant which does not require the cliques to be given, but instead searches for maximal cliques that satisfy the covering property.
    
    To see how, let $N_c$ be an upper bound on the number of cliques in the solution. Given $N_c$, let $c \in \set{1, \ldots, N_c}$ and instantiate the binary variables (i) $x_{v,c}$ where $v \in V(K)$, (ii) $y_{e,c}$ where $e \in E(K)$, and (iii) $z_c$. 
    Semantically, the variables in (i) will be used to track whether vertex $v$ (of $K$) is in clique $c$, those in (ii) to represent the edges in $c$ and the ones in (iii) stand for the active cliques in the solution. To find the relationships between these variables, first suppose that $z_c = 0$. Then obviously $x_{v,c}$ cannot be a vertex in the solution for any $v$, therefore $x_{v, c} \leq z_c$. Moreover, assume $e = \set{u,v} \in E(K)$. Then if a given clique $c$ does not contain vertex $v$, it cannot contain the edge $e$, leading to $y_{e,c} \leq x_{u,c}$ and $y_{e,c} \leq x_{v,c}$. On the other hand, if and both $x_{u,c} = x_{v,c} = 1$, then $y_{e,c}$ must exist, and to express that we can say $x_{u,c} + x_{v,c} - 1 \leq y_{e, c}$; but naturally, if $e \notin E(K)$, then $x_{u,c} + x_{v,c} \leq 1$. These relations constitute the ``clique finding'' part of the model, but we are yet to represent the edge covering constraints, which take a rather simple form: For each edge $e_G$ of the graph $G$ that we wish to cover, it must be true that $\sum_c y_{e_G, c} \geq 1$. Altogether, we have the following specification:
    \begin{subequations}\label{app-eq:mip-no-given-cliques}
    \begin{align}
        &\underset{\set{ z_c, x_{v, c}, y_{e, c} }}{\text{minimize: }} \;\;\;\sum_c z_c \\
        &\text{subject to: } \nonumber \\
        &\quad\;\;\; x_{v,c} \leq z_c \\
            &\quad\;\;\; \begin{rcases}
            y_{e, c} \leq x_{u, c} \\
            y_{e, c} \leq x_{v, c} \\
            x_{u,c} + x_{v,c} - 1 \leq y_{e, c} \\
            \end{rcases}
            &&\hspace{-1em}\text{if } e = \set{u, v} \in K \\
            &\quad\;\;\; x_{u,c} + x_{v, c} \leq 1 \;&&\hspace{-1em}\text{if } e = \set{u, v} \notin K\\
        &\quad\;\;\; \sum_c z_c y_{e, c} \geq 1 \;&&\hspace{-1em}\text{if } e \in G
    \end{align}
    \end{subequations}
    This is again a binary optimisation problem, and the number of variables is dependent on our initial guess $N_c$. Notice that in cases where no good guess $N_c$ is known and the user inputs a guess smaller than the optimal solution, the program will be infeasible.
    
    It is often the case in binary programming that whether a solution will be found or not strongly depends on the formulation of the model. Since both models above lead to optimal solutions, we have used them interchangeably depending on which of them performed better for each instance. In cases where the running time is too long, interrupting the process before it finds an optimal covering will return a valid, but potentially sub-optimal solution, as well as a lower-bound on the minimum number of necessary cliques. Such is the case of the data points with error bars in Fig.~\ref{fig:nof-settings-covering} of the main text.
    
    For the heuristic solutions we have employed a fairly straightforward heuristic algorithm modified from Ref.~\cite{conte_clique_2016}. In a nutshell, the idea is to (i) select an uncovered edge of $G$, (ii) grow a maximal clique in $K$ containing it, (iii) add it to the solution set and mark which other edges of $G$ were also covered by it, then (4) repeat until there are no more uncovered edges in $G$. 
    For the case where $K = G$, this algorithm was presented and its behaviour empirically analysed in Ref.~\cite{conte_clique_2016}. The case $K \neq G$ used here is a straightforward modification of it, thus we refer the interested reader to the original reference. 
    The quality of the solutions obtained with this method will crucially depend on how wasteful are the cliques grown in step (ii), and for this a heuristic to favour cliques covering a large number of uncovered edges of $G$ (as opposed to cliques that cover edges which are already contained in the solution) is used. Moreover, the edge selection in step (i) also has an effect on the quality of the solution. For this we have used used a random choice, ran the algorithm multiple times for each instance, and then selected the best run.
   
\medskip

\section{Measurements to obtain all four-point correlators}
\label{app:example-settings-four-point-kn}

Here we list the solutions found with the exact formulation for the four-point correlators problem with 3, 4 and 6 modes. These are of particular interest because, as discussed in the main text, they can be used as the basis for constructing the covering of triangular, rectangular and hexagonal lattices, respectively. Each column shown below corresponds to a measurement setting. For ease of reading, we use the same notation as in Fig.~\ref{fig:gm}, namely, the operators $S_{x}^{ij}$ and $S_{y}^{ij}$  are denoted as $X_{ij}$ and $Y_{ij}$, respectively.

\subsection{Three modes}
\label{app:2rdm-k3-settings}
\begin{equation*}
\begin{matrix}
1. & 2. & 3. & 4. & 5. & 6. & 7. \vspace{.5em}\\
n_1 & n_1 & n_1 & n_2 & n_2 & n_3 & n_3 \\
n_2 & X_{23} & Y_{23} & X_{13} & Y_{13} & Y_{12} & X_{12} \\
n_3 &  &  &  &  &  &  \\
\end{matrix}
\end{equation*}

\subsection{Four modes}
\label{app:2rdm-k4-settings}
\label{2_RDM_4}

\begin{equation*}
\begin{matrix}
1. & 2. & 3. & 4. & 5. & 6. & 7. & 8. & 9. & 10. \vspace{.5em}\\
n_1 & n_1 & n_1 & n_1 & n_1 & n_1 & n_2 & n_2 & n_2 & n_2  \\
n_2 & n_2 & n_3 & n_3 & n_4 & n_4 & n_3 & n_3 & n_4 & n_4 \\
X_{34} & Y_{34} & X_{24} & Y_{24} & X_{23} & Y_{23} & X_{14} & Y_{14} & X_{13} & Y_{13}  
\end{matrix}
\end{equation*} 
\begin{equation*}
\begin{matrix}
11. & 12. & 13. & 14. & 15. & 16. & 17. & 18. & 19. & 20. \vspace{.5em}\\
n_3 & n_3 & X_{12} & X_{12} & Y_{12} & Y_{12} & X_{13} & X_{13} & Y_{13} & Y_{13} \\
n_4 & n_4 & X_{34} & Y_{34} & X_{34} & Y_{34} & X_{24} & Y_{24} & X_{24} & Y_{24} \\
X_{12} & Y_{12} &  &  &  &  &  &  &  &  
\end{matrix}
\end{equation*}

\subsection{Six modes}
\label{app:2rdm-k6-settings}
\begin{equation*}
\begin{matrix}
    1. & 2. & 3. & 4. & 5. & 6. & 7. & 8. & 9. & 10. \vspace{.5em}\\
	n_1 & n_1 & n_1 & n_1 & n_1 & n_1 & n_1 & n_1 & n_1 & n_1 \\
	n_2 & n_2 & n_3 & n_3 & n_4 & n_4 & n_5 & n_5 & n_6 & n_6 \\
	Y_{34} & X_{36} & Y_{25} & Y_{26} & X_{25} & Y_{23} & X_{26} & X_{23} & X_{24} & X_{24} \\
	X_{56} & Y_{45} & X_{46} & X_{45} & Y_{36} & Y_{56} & X_{34} & Y_{46} & X_{35} & Y_{35} \\
\end{matrix}
\end{equation*}

\begin{equation*}
\begin{matrix}
    11. & 12. & 13. & 14. & 15. & 16. & 17. & 18. & 19. & 20. \vspace{.5em}\\
	n_1 & n_1 & n_1 & n_1 & n_1 & n_1 & n_2 & n_2 & n_2 & n_2 \\ 
	n_6 & n_6 & n_6 & n_6 & n_6 & n_6 & n_3 & n_3 & n_4 & n_4 \\
	Y_{24} & Y_{24} & X_{23} & X_{23} & Y_{23} & Y_{23} & Y_{14} & Y_{15} & X_{16} & Y_{16} \\  
	X_{35} & Y_{35} & X_{45} & Y_{45} & X_{45} & Y_{45} & Y_{56} & Y_{46} & X_{35} & Y_{35} \\  
\end{matrix}
\end{equation*}

\begin{equation*}
\begin{matrix}
    21. & 22. & 23. & 24. & 25. & 26. & 27. & 28. & 29. & 30. \vspace{.5em}\\
	n_2 & n_2 & n_2 & n_2 & n_3 & n_3 & n_3 & n_3 & n_3 & n_3 \\
	n_5 & n_5 & n_6 & n_6 & n_5 & n_5 & n_4 & n_4 & n_4 & n_4 \\
	Y_{13} & X_{14} & X_{13} & X_{15} & X_{16} & Y_{16} & X_{12} & X_{15} & X_{15} & Y_{15} \\
	X_{46} & Y_{36} & X_{45} & X_{34} & X_{24} & Y_{24} & X_{56} & X_{26} & Y_{26} & X_{26} \\
\end{matrix}
\end{equation*}

\begin{equation*}
\begin{matrix}
    31. & 32. & 33. & 34. & 35. & 36. & 37. & 38. & 39. & 40. \vspace{.5em}\\
	n_3 & n_3 & n_3 & n_4 & n_4 & n_4 & n_4 & n_5 & n_5 & X_{12} \\
	n_4 & n_6 & n_6 & n_5 & n_5 & n_6 & n_6 & n_6 & n_6 & X_{35} \\ 
	Y_{15} & Y_{12} & X_{14} & Y_{12} & X_{13} & y_{13} & Y_{15} & X_{12} & Y_{14} & Y_{46} \\
	Y_{26} & Y_{45} & X_{25} & X_{36} & Y_{26} & y_{25} & X_{23} & Y_{34} & Y_{23} \\ 
\end{matrix}
\end{equation*}

\begin{equation*}
\begin{matrix}
    41. & 42. & 43. & 44. & 45. & 46. & 47. & 48. & 49. & 50. \vspace{.5em}\\
	X_{12} & Y_{12} & Y_{12} & X_{12} & X_{12} & Y_{12} & X_{12} & Y_{12} & Y_{12} & X_{13} \\ 
	Y_{35} & X_{35} & Y_{35} & X_{45} & Y_{45} & X_{45} & Y_{56} & X_{56} & Y_{56} & X_{25} \\ 
	X_{46} & X_{46} & Y_{46} & Y_{36} & X_{36} & Y_{36} & X_{34} & X_{34} & Y_{34} & X_{46} \\ 
\end{matrix}
\end{equation*}

\begin{equation*}
\begin{matrix}
    51. & 52. & 53. & 54. & 55. & 56. & 57. & 58. & 59. & 60. \vspace{.5em}\\
	X_{13} & Y_{13} & X_{13} & Y_{13} & Y_{13} & X_{13} & X_{13} & Y_{13} & Y_{13} & X_{14} \\ 
	Y_{25} & X_{25} & X_{26} & X_{26} & Y_{26} & X_{24} & Y_{24} & X_{24} & Y_{24} & X_{23} \\ 
	Y_{46} & Y_{46} & Y_{45} & X_{45} & Y_{45} & Y_{56} & X_{56} & X_{56} & Y_{56} & X_{56} \\ 
\end{matrix}
\end{equation*}

\begin{equation*}
\begin{matrix}
    61. & 62. & 63. & 64. & 65. & 66. & 67. & 68. & 69. & 70. \vspace{.5em}\\
	X_{14} & Y_{14} & X_{14} & Y_{14} & Y_{14} & X_{14} & X_{14} & Y_{14} & Y_{14} & X_{15} \\ 
	X_{23} & Y_{23} & Y_{25} & X_{25} & Y_{25} & X_{26} & Y_{26} & X_{26} & Y_{26} & X_{24} \\ 
	Y_{56} & X_{56} & X_{36} & X_{36} & Y_{36} & X_{35} & Y_{35} & X_{35} & Y_{35} & X_{36} \\ 
\end{matrix}
\end{equation*}

\begin{equation*}
\begin{matrix}
    71. & 72. & 73. & 74. & 75. & 76. \vspace{.5em}\\
	X_{15} & Y_{15} & Y_{15} & X_{15} & X_{15} & Y_{15} \\ 
	X_{24} & Y_{24} & Y_{24} & Y_{23} & Y_{23} & X_{23} \\ 
	Y_{36} & X_{36} & Y_{36} & Y_{46} & X_{46} & X_{46} \\ 
\end{matrix}
\end{equation*}

\end{appendix}
\clearpage

\bibliographystyle{apsrev4-2}
\bibliography{bibliography}

\begin{thebibliography}{51}%
\makeatletter
\providecommand \@ifxundefined [1]{%
 \@ifx{#1\undefined}
}%
\providecommand \@ifnum [1]{%
 \ifnum #1\expandafter \@firstoftwo
 \else \expandafter \@secondoftwo
 \fi
}%
\providecommand \@ifx [1]{%
 \ifx #1\expandafter \@firstoftwo
 \else \expandafter \@secondoftwo
 \fi
}%
\providecommand \natexlab [1]{#1}%
\providecommand \enquote  [1]{``#1''}%
\providecommand \bibnamefont  [1]{#1}%
\providecommand \bibfnamefont [1]{#1}%
\providecommand \citenamefont [1]{#1}%
\providecommand \href@noop [0]{\@secondoftwo}%
\providecommand \href [0]{\begingroup \@sanitize@url \@href}%
\providecommand \@href[1]{\@@startlink{#1}\@@href}%
\providecommand \@@href[1]{\endgroup#1\@@endlink}%
\providecommand \@sanitize@url [0]{\catcode `\\12\catcode `\$12\catcode `\&12\catcode `\#12\catcode `\^12\catcode `\_12\catcode `\%12\relax}%
\providecommand \@@startlink[1]{}%
\providecommand \@@endlink[0]{}%
\providecommand \url  [0]{\begingroup\@sanitize@url \@url }%
\providecommand \@url [1]{\endgroup\@href {#1}{\urlprefix }}%
\providecommand \urlprefix  [0]{URL }%
\providecommand \Eprint [0]{\href }%
\providecommand \doibase [0]{https://doi.org/}%
\providecommand \selectlanguage [0]{\@gobble}%
\providecommand \bibinfo  [0]{\@secondoftwo}%
\providecommand \bibfield  [0]{\@secondoftwo}%
\providecommand \translation [1]{[#1]}%
\providecommand \BibitemOpen [0]{}%
\providecommand \bibitemStop [0]{}%
\providecommand \bibitemNoStop [0]{.\EOS\space}%
\providecommand \EOS [0]{\spacefactor3000\relax}%
\providecommand \BibitemShut  [1]{\csname bibitem#1\endcsname}%
\let\auto@bib@innerbib\@empty
\bibitem [{\citenamefont {Georgescu}\ \emph {et~al.}(2014)\citenamefont {Georgescu}, \citenamefont {Ashhab},\ and\ \citenamefont {Nori}}]{Georgescu2014}%
  \BibitemOpen
  \bibfield  {author} {\bibinfo {author} {\bibfnamefont {I.~M.}\ \bibnamefont {Georgescu}}, \bibinfo {author} {\bibfnamefont {S.}~\bibnamefont {Ashhab}},\ and\ \bibinfo {author} {\bibfnamefont {F.}~\bibnamefont {Nori}},\ }\href {https://doi.org/10.1103/RevModPhys.86.153} {\bibfield  {journal} {\bibinfo  {journal} {Rev. Mod. Phys.}\ }\textbf {\bibinfo {volume} {86}},\ \bibinfo {pages} {153} (\bibinfo {year} {2014})}\BibitemShut {NoStop}%
\bibitem [{\citenamefont {Altman}\ \emph {et~al.}(2021)\citenamefont {Altman}, \citenamefont {Brown}, \citenamefont {Carleo}, \citenamefont {Carr}, \citenamefont {Demler} \emph {et~al.}}]{Altman2021}%
  \BibitemOpen
  \bibfield  {author} {\bibinfo {author} {\bibfnamefont {E.}~\bibnamefont {Altman}}, \bibinfo {author} {\bibfnamefont {K.~R.}\ \bibnamefont {Brown}}, \bibinfo {author} {\bibfnamefont {G.}~\bibnamefont {Carleo}}, \bibinfo {author} {\bibfnamefont {L.~D.}\ \bibnamefont {Carr}}, \bibinfo {author} {\bibfnamefont {E.}~\bibnamefont {Demler}}, \emph {et~al.},\ }\href {https://doi.org/10.1103/PRXQuantum.2.017003} {\bibfield  {journal} {\bibinfo  {journal} {PRX Quantum}\ }\textbf {\bibinfo {volume} {2}},\ \bibinfo {pages} {017003} (\bibinfo {year} {2021})}\BibitemShut {NoStop}%
\bibitem [{\citenamefont {Paudel}\ \emph {et~al.}(2022)\citenamefont {Paudel}, \citenamefont {Syamlal}, \citenamefont {Crawford}, \citenamefont {Lee}, \citenamefont {Shugayev} \emph {et~al.}}]{Paudel2022}%
  \BibitemOpen
  \bibfield  {author} {\bibinfo {author} {\bibfnamefont {H.~P.}\ \bibnamefont {Paudel}}, \bibinfo {author} {\bibfnamefont {M.}~\bibnamefont {Syamlal}}, \bibinfo {author} {\bibfnamefont {S.~E.}\ \bibnamefont {Crawford}}, \bibinfo {author} {\bibfnamefont {Y.-L.}\ \bibnamefont {Lee}}, \bibinfo {author} {\bibfnamefont {R.~A.}\ \bibnamefont {Shugayev}}, \emph {et~al.},\ }\href {https://doi.org/10.1021/acsengineeringau.1c00033} {\bibfield  {journal} {\bibinfo  {journal} {ACS Eng. Au}\ }\textbf {\bibinfo {volume} {2}},\ \bibinfo {pages} {151} (\bibinfo {year} {2022})}\BibitemShut {NoStop}%
\bibitem [{\citenamefont {Fraxanet}\ \emph {et~al.}(2023)\citenamefont {Fraxanet}, \citenamefont {Salamon},\ and\ \citenamefont {Lewenstein}}]{Fraxanet2023}%
  \BibitemOpen
  \bibfield  {author} {\bibinfo {author} {\bibfnamefont {J.}~\bibnamefont {Fraxanet}}, \bibinfo {author} {\bibfnamefont {T.}~\bibnamefont {Salamon}},\ and\ \bibinfo {author} {\bibfnamefont {M.}~\bibnamefont {Lewenstein}},\ }\bibinfo {title} {The coming decades of quantum simulation},\ in\ \href {https://doi.org/10.1007/978-3-031-32469-7_4} {\emph {\bibinfo {booktitle} {Sketches of Physics: The Celebration Collection}}},\ \bibinfo {editor} {edited by\ \bibinfo {editor} {\bibfnamefont {R.}~\bibnamefont {Citro}}, \bibinfo {editor} {\bibfnamefont {M.}~\bibnamefont {Lewenstein}}, \bibinfo {editor} {\bibfnamefont {A.}~\bibnamefont {Rubio}}, \bibinfo {editor} {\bibfnamefont {W.~P.}\ \bibnamefont {Schleich}}, \bibinfo {editor} {\bibfnamefont {J.~D.}\ \bibnamefont {Wells}},\ and\ \bibinfo {editor} {\bibfnamefont {G.~P.}\ \bibnamefont {Zank}}}\ (\bibinfo  {publisher} {Springer International Publishing},\ \bibinfo {address} {Cham},\ \bibinfo {year} {2023})\ pp.\ \bibinfo {pages} {85--125}\BibitemShut {NoStop}%
\bibitem [{\citenamefont {Arute}\ \emph {et~al.}(2019)\citenamefont {Arute}, \citenamefont {Arya}, \citenamefont {Babbush}, \citenamefont {Bacon}, \citenamefont {Bardin} \emph {et~al.}}]{Arute2019}%
  \BibitemOpen
  \bibfield  {author} {\bibinfo {author} {\bibfnamefont {F.}~\bibnamefont {Arute}}, \bibinfo {author} {\bibfnamefont {K.}~\bibnamefont {Arya}}, \bibinfo {author} {\bibfnamefont {R.}~\bibnamefont {Babbush}}, \bibinfo {author} {\bibfnamefont {D.}~\bibnamefont {Bacon}}, \bibinfo {author} {\bibfnamefont {J.~C.}\ \bibnamefont {Bardin}}, \emph {et~al.},\ }\href {https://doi.org/10.1038/s41586-019-1666-5} {\bibfield  {journal} {\bibinfo  {journal} {Nature}\ }\textbf {\bibinfo {volume} {574}},\ \bibinfo {pages} {505–510} (\bibinfo {year} {2019})}\BibitemShut {NoStop}%
\bibitem [{\citenamefont {Bluvstein}\ \emph {et~al.}(2023)\citenamefont {Bluvstein}, \citenamefont {Evered}, \citenamefont {Geim}, \citenamefont {Li}, \citenamefont {Zhou} \emph {et~al.}}]{Bluvstein2023}%
  \BibitemOpen
  \bibfield  {author} {\bibinfo {author} {\bibfnamefont {D.}~\bibnamefont {Bluvstein}}, \bibinfo {author} {\bibfnamefont {S.~J.}\ \bibnamefont {Evered}}, \bibinfo {author} {\bibfnamefont {A.~A.}\ \bibnamefont {Geim}}, \bibinfo {author} {\bibfnamefont {S.~H.}\ \bibnamefont {Li}}, \bibinfo {author} {\bibfnamefont {H.}~\bibnamefont {Zhou}}, \emph {et~al.},\ }\href {https://doi.org/10.1038/s41586-023-06927-3} {\bibfield  {journal} {\bibinfo  {journal} {Nature}\ }\textbf {\bibinfo {volume} {626}},\ \bibinfo {pages} {58–65} (\bibinfo {year} {2023})}\BibitemShut {NoStop}%
\bibitem [{\citenamefont {Kim}\ \emph {et~al.}(2023)\citenamefont {Kim}, \citenamefont {Eddins}, \citenamefont {Anand}, \citenamefont {Wei}, \citenamefont {van~den Berg} \emph {et~al.}}]{Kim2023}%
  \BibitemOpen
  \bibfield  {author} {\bibinfo {author} {\bibfnamefont {Y.}~\bibnamefont {Kim}}, \bibinfo {author} {\bibfnamefont {A.}~\bibnamefont {Eddins}}, \bibinfo {author} {\bibfnamefont {S.}~\bibnamefont {Anand}}, \bibinfo {author} {\bibfnamefont {K.~X.}\ \bibnamefont {Wei}}, \bibinfo {author} {\bibfnamefont {E.}~\bibnamefont {van~den Berg}}, \emph {et~al.},\ }\href {https://doi.org/10.1038/s41586-023-06096-3} {\bibfield  {journal} {\bibinfo  {journal} {Nature}\ }\textbf {\bibinfo {volume} {618}},\ \bibinfo {pages} {500–505} (\bibinfo {year} {2023})}\BibitemShut {NoStop}%
\bibitem [{\citenamefont {Acharya}\ \emph {et~al.}(2023)\citenamefont {Acharya}, \citenamefont {Aleiner}, \citenamefont {Allen}, \citenamefont {Andersen}, \citenamefont {Ansmann} \emph {et~al.}}]{googleai2023}%
  \BibitemOpen
  \bibfield  {author} {\bibinfo {author} {\bibfnamefont {R.}~\bibnamefont {Acharya}}, \bibinfo {author} {\bibfnamefont {I.}~\bibnamefont {Aleiner}}, \bibinfo {author} {\bibfnamefont {R.}~\bibnamefont {Allen}}, \bibinfo {author} {\bibfnamefont {T.~I.}\ \bibnamefont {Andersen}}, \bibinfo {author} {\bibfnamefont {M.}~\bibnamefont {Ansmann}}, \emph {et~al.},\ }\href {https://doi.org/10.1038/s41586-022-05434-1} {\bibfield  {journal} {\bibinfo  {journal} {Nature}\ }\textbf {\bibinfo {volume} {614}},\ \bibinfo {pages} {676–681} (\bibinfo {year} {2023})}\BibitemShut {NoStop}%
\bibitem [{\citenamefont {Jordan}\ and\ \citenamefont {Wigner}(1928)}]{Jordan1928}%
  \BibitemOpen
  \bibfield  {author} {\bibinfo {author} {\bibfnamefont {P.}~\bibnamefont {Jordan}}\ and\ \bibinfo {author} {\bibfnamefont {E.}~\bibnamefont {Wigner}},\ }\href {https://doi.org/10.1007/BF01331938} {\bibfield  {journal} {\bibinfo  {journal} {Z. Phys.}\ }\textbf {\bibinfo {volume} {47}},\ \bibinfo {pages} {631} (\bibinfo {year} {1928})}\BibitemShut {NoStop}%
\bibitem [{\citenamefont {Bravyi}\ and\ \citenamefont {Kitaev}(2002)}]{Bravi2002}%
  \BibitemOpen
  \bibfield  {author} {\bibinfo {author} {\bibfnamefont {S.~B.}\ \bibnamefont {Bravyi}}\ and\ \bibinfo {author} {\bibfnamefont {A.~Y.}\ \bibnamefont {Kitaev}},\ }\href {https://doi.org/https://doi.org/10.1006/aphy.2002.6254} {\bibfield  {journal} {\bibinfo  {journal} {Ann. Phys.}\ }\textbf {\bibinfo {volume} {298}},\ \bibinfo {pages} {210} (\bibinfo {year} {2002})}\BibitemShut {NoStop}%
\bibitem [{\citenamefont {Jiang}\ \emph {et~al.}(2020)\citenamefont {Jiang}, \citenamefont {Kalev}, \citenamefont {Mruczkiewicz},\ and\ \citenamefont {Neven}}]{Jiang2020}%
  \BibitemOpen
  \bibfield  {author} {\bibinfo {author} {\bibfnamefont {Z.}~\bibnamefont {Jiang}}, \bibinfo {author} {\bibfnamefont {A.}~\bibnamefont {Kalev}}, \bibinfo {author} {\bibfnamefont {W.}~\bibnamefont {Mruczkiewicz}},\ and\ \bibinfo {author} {\bibfnamefont {H.}~\bibnamefont {Neven}},\ }\href {https://doi.org/10.22331/q-2020-06-04-276} {\bibfield  {journal} {\bibinfo  {journal} {{Quantum}}\ }\textbf {\bibinfo {volume} {4}},\ \bibinfo {pages} {276} (\bibinfo {year} {2020})}\BibitemShut {NoStop}%
\bibitem [{\citenamefont {Norcia}\ \emph {et~al.}(2018)\citenamefont {Norcia}, \citenamefont {Young},\ and\ \citenamefont {Kaufman}}]{Norcia2018}%
  \BibitemOpen
  \bibfield  {author} {\bibinfo {author} {\bibfnamefont {M.~A.}\ \bibnamefont {Norcia}}, \bibinfo {author} {\bibfnamefont {A.~W.}\ \bibnamefont {Young}},\ and\ \bibinfo {author} {\bibfnamefont {A.~M.}\ \bibnamefont {Kaufman}},\ }\href {https://doi.org/10.1103/PhysRevX.8.041054} {\bibfield  {journal} {\bibinfo  {journal} {Phys. Rev. X}\ }\textbf {\bibinfo {volume} {8}},\ \bibinfo {pages} {041054} (\bibinfo {year} {2018})}\BibitemShut {NoStop}%
\bibitem [{\citenamefont {Yan}\ \emph {et~al.}(2022)\citenamefont {Yan}, \citenamefont {Spar}, \citenamefont {Prichard}, \citenamefont {Chi}, \citenamefont {Wei} \emph {et~al.}}]{Yan2022}%
  \BibitemOpen
  \bibfield  {author} {\bibinfo {author} {\bibfnamefont {Z.~Z.}\ \bibnamefont {Yan}}, \bibinfo {author} {\bibfnamefont {B.~M.}\ \bibnamefont {Spar}}, \bibinfo {author} {\bibfnamefont {M.~L.}\ \bibnamefont {Prichard}}, \bibinfo {author} {\bibfnamefont {S.}~\bibnamefont {Chi}}, \bibinfo {author} {\bibfnamefont {H.-T.}\ \bibnamefont {Wei}}, \emph {et~al.},\ }\href {https://doi.org/10.1103/PhysRevLett.129.123201} {\bibfield  {journal} {\bibinfo  {journal} {Phys. Rev. Lett.}\ }\textbf {\bibinfo {volume} {129}},\ \bibinfo {pages} {123201} (\bibinfo {year} {2022})}\BibitemShut {NoStop}%
\bibitem [{\citenamefont {Spar}\ \emph {et~al.}(2022)\citenamefont {Spar}, \citenamefont {Guardado-Sanchez}, \citenamefont {Chi}, \citenamefont {Yan},\ and\ \citenamefont {Bakr}}]{Spar2022}%
  \BibitemOpen
  \bibfield  {author} {\bibinfo {author} {\bibfnamefont {B.~M.}\ \bibnamefont {Spar}}, \bibinfo {author} {\bibfnamefont {E.}~\bibnamefont {Guardado-Sanchez}}, \bibinfo {author} {\bibfnamefont {S.}~\bibnamefont {Chi}}, \bibinfo {author} {\bibfnamefont {Z.~Z.}\ \bibnamefont {Yan}},\ and\ \bibinfo {author} {\bibfnamefont {W.~S.}\ \bibnamefont {Bakr}},\ }\href {https://doi.org/10.1103/PhysRevLett.128.223202} {\bibfield  {journal} {\bibinfo  {journal} {Phys. Rev. Lett.}\ }\textbf {\bibinfo {volume} {128}},\ \bibinfo {pages} {223202} (\bibinfo {year} {2022})}\BibitemShut {NoStop}%
\bibitem [{\citenamefont {González-Cuadra}\ \emph {et~al.}(2023)\citenamefont {González-Cuadra}, \citenamefont {Bluvstein}, \citenamefont {Kalinowski}, \citenamefont {Kaubruegger}, \citenamefont {Maskara} \emph {et~al.}}]{González-Cuadra2023}%
  \BibitemOpen
  \bibfield  {author} {\bibinfo {author} {\bibfnamefont {D.}~\bibnamefont {González-Cuadra}}, \bibinfo {author} {\bibfnamefont {D.}~\bibnamefont {Bluvstein}}, \bibinfo {author} {\bibfnamefont {M.}~\bibnamefont {Kalinowski}}, \bibinfo {author} {\bibfnamefont {R.}~\bibnamefont {Kaubruegger}}, \bibinfo {author} {\bibfnamefont {N.}~\bibnamefont {Maskara}}, \emph {et~al.},\ }\href {https://doi.org/10.1073/pnas.2304294120} {\bibfield  {journal} {\bibinfo  {journal} {PNAS}\ }\textbf {\bibinfo {volume} {120}},\ \bibinfo {pages} {e2304294120} (\bibinfo {year} {2023})}\BibitemShut {NoStop}%
\bibitem [{\citenamefont {Arg{\"u}ello-Luengo}\ \emph {et~al.}(2019)\citenamefont {Arg{\"u}ello-Luengo}, \citenamefont {Gonz{\'a}lez-Tudela}, \citenamefont {Shi}, \citenamefont {Zoller},\ and\ \citenamefont {Cirac}}]{Argüello-Luengo2019}%
  \BibitemOpen
  \bibfield  {author} {\bibinfo {author} {\bibfnamefont {J.}~\bibnamefont {Arg{\"u}ello-Luengo}}, \bibinfo {author} {\bibfnamefont {A.}~\bibnamefont {Gonz{\'a}lez-Tudela}}, \bibinfo {author} {\bibfnamefont {T.}~\bibnamefont {Shi}}, \bibinfo {author} {\bibfnamefont {P.}~\bibnamefont {Zoller}},\ and\ \bibinfo {author} {\bibfnamefont {J.~I.}\ \bibnamefont {Cirac}},\ }\href {https://doi.org/10.1038/s41586-019-1614-4} {\bibfield  {journal} {\bibinfo  {journal} {Nature}\ }\textbf {\bibinfo {volume} {574}},\ \bibinfo {pages} {215} (\bibinfo {year} {2019})}\BibitemShut {NoStop}%
\bibitem [{\citenamefont {Schuckert}\ \emph {et~al.}()\citenamefont {Schuckert}, \citenamefont {Crane}, \citenamefont {Gorshkov}, \citenamefont {Hafezi},\ and\ \citenamefont {Gullans}}]{schuckert2024}%
  \BibitemOpen
  \bibfield  {author} {\bibinfo {author} {\bibfnamefont {A.}~\bibnamefont {Schuckert}}, \bibinfo {author} {\bibfnamefont {E.}~\bibnamefont {Crane}}, \bibinfo {author} {\bibfnamefont {A.~V.}\ \bibnamefont {Gorshkov}}, \bibinfo {author} {\bibfnamefont {M.}~\bibnamefont {Hafezi}},\ and\ \bibinfo {author} {\bibfnamefont {M.~J.}\ \bibnamefont {Gullans}},\ }\href@noop {} {}\Eprint {https://arxiv.org/abs/2411.08955v2} {arXiv:2411.08955v2 [quant-ph]} \BibitemShut {NoStop}%
\bibitem [{\citenamefont {Tabares}\ \emph {et~al.}(2025)\citenamefont {Tabares}, \citenamefont {Kokail}, \citenamefont {Zoller}, \citenamefont {Gonz\'alez-Cuadra},\ and\ \citenamefont {Gonz\'alez-Tudela}}]{tabares2025}%
  \BibitemOpen
  \bibfield  {author} {\bibinfo {author} {\bibfnamefont {C.}~\bibnamefont {Tabares}}, \bibinfo {author} {\bibfnamefont {C.}~\bibnamefont {Kokail}}, \bibinfo {author} {\bibfnamefont {P.}~\bibnamefont {Zoller}}, \bibinfo {author} {\bibfnamefont {D.}~\bibnamefont {Gonz\'alez-Cuadra}},\ and\ \bibinfo {author} {\bibfnamefont {A.}~\bibnamefont {Gonz\'alez-Tudela}},\ }\href {https://doi.org/10.1103/3nx4-bnyy} {\bibfield  {journal} {\bibinfo  {journal} {PRX Quantum}\ }\textbf {\bibinfo {volume} {6}},\ \bibinfo {pages} {030356} (\bibinfo {year} {2025})}\BibitemShut {NoStop}%
\bibitem [{\citenamefont {Gkritsis}\ \emph {et~al.}(2025)\citenamefont {Gkritsis}, \citenamefont {Dux}, \citenamefont {Zhang}, \citenamefont {Jain}, \citenamefont {Gogolin},\ and\ \citenamefont {Preiss}}]{gkritsis2025simulating}%
  \BibitemOpen
  \bibfield  {author} {\bibinfo {author} {\bibfnamefont {F.}~\bibnamefont {Gkritsis}}, \bibinfo {author} {\bibfnamefont {D.}~\bibnamefont {Dux}}, \bibinfo {author} {\bibfnamefont {J.}~\bibnamefont {Zhang}}, \bibinfo {author} {\bibfnamefont {N.}~\bibnamefont {Jain}}, \bibinfo {author} {\bibfnamefont {C.}~\bibnamefont {Gogolin}},\ and\ \bibinfo {author} {\bibfnamefont {P.~M.}\ \bibnamefont {Preiss}},\ }\href {https://doi.org/10.1103/PRXQuantum.6.010318} {\bibfield  {journal} {\bibinfo  {journal} {PRX Quantum}\ }\textbf {\bibinfo {volume} {6}},\ \bibinfo {pages} {010318} (\bibinfo {year} {2025})}\BibitemShut {NoStop}%
\bibitem [{\citenamefont {Henriet}\ \emph {et~al.}(2020)\citenamefont {Henriet}, \citenamefont {Beguin}, \citenamefont {Signoles}, \citenamefont {Lahaye}, \citenamefont {Browaeys} \emph {et~al.}}]{Henriet2020}%
  \BibitemOpen
  \bibfield  {author} {\bibinfo {author} {\bibfnamefont {L.}~\bibnamefont {Henriet}}, \bibinfo {author} {\bibfnamefont {L.}~\bibnamefont {Beguin}}, \bibinfo {author} {\bibfnamefont {A.}~\bibnamefont {Signoles}}, \bibinfo {author} {\bibfnamefont {T.}~\bibnamefont {Lahaye}}, \bibinfo {author} {\bibfnamefont {A.}~\bibnamefont {Browaeys}}, \emph {et~al.},\ }\href {https://doi.org/10.22331/q-2020-09-21-327} {\bibfield  {journal} {\bibinfo  {journal} {{Quantum}}\ }\textbf {\bibinfo {volume} {4}},\ \bibinfo {pages} {327} (\bibinfo {year} {2020})}\BibitemShut {NoStop}%
\bibitem [{\citenamefont {Evered}\ \emph {et~al.}(2023)\citenamefont {Evered}, \citenamefont {Bluvstein}, \citenamefont {Kalinowski}, \citenamefont {Ebadi}, \citenamefont {Manovitz} \emph {et~al.}}]{Evered2023}%
  \BibitemOpen
  \bibfield  {author} {\bibinfo {author} {\bibfnamefont {S.~J.}\ \bibnamefont {Evered}}, \bibinfo {author} {\bibfnamefont {D.}~\bibnamefont {Bluvstein}}, \bibinfo {author} {\bibfnamefont {M.}~\bibnamefont {Kalinowski}}, \bibinfo {author} {\bibfnamefont {S.}~\bibnamefont {Ebadi}}, \bibinfo {author} {\bibfnamefont {T.}~\bibnamefont {Manovitz}}, \emph {et~al.},\ }\href {https://doi.org/10.1038/s41586-023-06481-y} {\bibfield  {journal} {\bibinfo  {journal} {Nature}\ }\textbf {\bibinfo {volume} {622}},\ \bibinfo {pages} {268} (\bibinfo {year} {2023})}\BibitemShut {NoStop}%
\bibitem [{\citenamefont {Wintersperger}\ \emph {et~al.}(2023)\citenamefont {Wintersperger}, \citenamefont {Dommert}, \citenamefont {Ehmer}, \citenamefont {Hoursanov}, \citenamefont {Klepsch} \emph {et~al.}}]{Wintersperger2023}%
  \BibitemOpen
  \bibfield  {author} {\bibinfo {author} {\bibfnamefont {K.}~\bibnamefont {Wintersperger}}, \bibinfo {author} {\bibfnamefont {F.}~\bibnamefont {Dommert}}, \bibinfo {author} {\bibfnamefont {T.}~\bibnamefont {Ehmer}}, \bibinfo {author} {\bibfnamefont {A.}~\bibnamefont {Hoursanov}}, \bibinfo {author} {\bibfnamefont {J.}~\bibnamefont {Klepsch}}, \emph {et~al.},\ }\href {https://doi.org/10.1140/epjqt/s40507-023-00190-1} {\bibfield  {journal} {\bibinfo  {journal} {EPJ Quantum Technol.}\ }\textbf {\bibinfo {volume} {10}},\ \bibinfo {pages} {32} (\bibinfo {year} {2023})}\BibitemShut {NoStop}%
\bibitem [{\citenamefont {Li}\ \emph {et~al.}(2023)\citenamefont {Li}, \citenamefont {Mukhopadhyay},\ and\ \citenamefont {Bayat}}]{Li2023}%
  \BibitemOpen
  \bibfield  {author} {\bibinfo {author} {\bibfnamefont {Q.}~\bibnamefont {Li}}, \bibinfo {author} {\bibfnamefont {C.}~\bibnamefont {Mukhopadhyay}},\ and\ \bibinfo {author} {\bibfnamefont {A.}~\bibnamefont {Bayat}},\ }\href {https://doi.org/10.1103/PhysRevResearch.5.043175} {\bibfield  {journal} {\bibinfo  {journal} {Phys. Rev. Res.}\ }\textbf {\bibinfo {volume} {5}},\ \bibinfo {pages} {043175} (\bibinfo {year} {2023})}\BibitemShut {NoStop}%
\bibitem [{\citenamefont {Bonet-Monroig}\ \emph {et~al.}(2020)\citenamefont {Bonet-Monroig}, \citenamefont {Babbush},\ and\ \citenamefont {O'Brien}}]{Monroig2020}%
  \BibitemOpen
  \bibfield  {author} {\bibinfo {author} {\bibfnamefont {X.}~\bibnamefont {Bonet-Monroig}}, \bibinfo {author} {\bibfnamefont {R.}~\bibnamefont {Babbush}},\ and\ \bibinfo {author} {\bibfnamefont {T.~E.}\ \bibnamefont {O'Brien}},\ }\href {https://doi.org/10.1103/PhysRevX.10.031064} {\bibfield  {journal} {\bibinfo  {journal} {Phys. Rev. X}\ }\textbf {\bibinfo {volume} {10}},\ \bibinfo {pages} {031064} (\bibinfo {year} {2020})}\BibitemShut {NoStop}%
\bibitem [{\citenamefont {Ara{\'u}jo}\ \emph {et~al.}(2022)\citenamefont {Ara{\'u}jo}, \citenamefont {Taddei}, \citenamefont {Cavalcanti},\ and\ \citenamefont {Ac{\'\i}n}}]{araujo2022local}%
  \BibitemOpen
  \bibfield  {author} {\bibinfo {author} {\bibfnamefont {B.~G.~M.}\ \bibnamefont {Ara{\'u}jo}}, \bibinfo {author} {\bibfnamefont {M.~M.}\ \bibnamefont {Taddei}}, \bibinfo {author} {\bibfnamefont {D.}~\bibnamefont {Cavalcanti}},\ and\ \bibinfo {author} {\bibfnamefont {A.}~\bibnamefont {Ac{\'\i}n}},\ }\href {https://doi.org/10.1103/PhysRevA.106.062441} {\bibfield  {journal} {\bibinfo  {journal} {Phys. Rev. A}\ }\textbf {\bibinfo {volume} {106}},\ \bibinfo {pages} {062441} (\bibinfo {year} {2022})}\BibitemShut {NoStop}%
\bibitem [{\citenamefont {Cotler}\ and\ \citenamefont {Wilczek}(2020)}]{Cotler2020}%
  \BibitemOpen
  \bibfield  {author} {\bibinfo {author} {\bibfnamefont {J.}~\bibnamefont {Cotler}}\ and\ \bibinfo {author} {\bibfnamefont {F.}~\bibnamefont {Wilczek}},\ }\href {https://doi.org/10.1103/PhysRevLett.124.100401} {\bibfield  {journal} {\bibinfo  {journal} {Phys. Rev. Lett.}\ }\textbf {\bibinfo {volume} {124}},\ \bibinfo {pages} {100401} (\bibinfo {year} {2020})}\BibitemShut {NoStop}%
\bibitem [{\citenamefont {Garc\'{\i}a-P\'erez}\ \emph {et~al.}(2020)\citenamefont {Garc\'{\i}a-P\'erez}, \citenamefont {Rossi}, \citenamefont {Sokolov}, \citenamefont {Borrelli},\ and\ \citenamefont {Maniscalco}}]{Maniscalco2020}%
  \BibitemOpen
  \bibfield  {author} {\bibinfo {author} {\bibfnamefont {G.}~\bibnamefont {Garc\'{\i}a-P\'erez}}, \bibinfo {author} {\bibfnamefont {M.~A.~C.}\ \bibnamefont {Rossi}}, \bibinfo {author} {\bibfnamefont {B.}~\bibnamefont {Sokolov}}, \bibinfo {author} {\bibfnamefont {E.-M.}\ \bibnamefont {Borrelli}},\ and\ \bibinfo {author} {\bibfnamefont {S.}~\bibnamefont {Maniscalco}},\ }\href {https://doi.org/10.1103/PhysRevResearch.2.023393} {\bibfield  {journal} {\bibinfo  {journal} {Phys. Rev. Res.}\ }\textbf {\bibinfo {volume} {2}},\ \bibinfo {pages} {023393} (\bibinfo {year} {2020})}\BibitemShut {NoStop}%
\bibitem [{\citenamefont {Verteletskyi}\ \emph {et~al.}(2020)\citenamefont {Verteletskyi}, \citenamefont {Yen},\ and\ \citenamefont {Izmaylov}}]{verteletskyi2020measurement}%
  \BibitemOpen
  \bibfield  {author} {\bibinfo {author} {\bibfnamefont {V.}~\bibnamefont {Verteletskyi}}, \bibinfo {author} {\bibfnamefont {T.-C.}\ \bibnamefont {Yen}},\ and\ \bibinfo {author} {\bibfnamefont {A.~F.}\ \bibnamefont {Izmaylov}},\ }\href {https://doi.org/10.1063/1.5141458} {\bibfield  {journal} {\bibinfo  {journal} {J. Chem. Phys.}\ }\textbf {\bibinfo {volume} {152}},\ \bibinfo {pages} {124114} (\bibinfo {year} {2020})}\BibitemShut {NoStop}%
\bibitem [{\citenamefont {Jaloveckas}\ \emph {et~al.}(2023)\citenamefont {Jaloveckas}, \citenamefont {Nguyen}, \citenamefont {Palackal}, \citenamefont {Lorenz},\ and\ \citenamefont {Ehm}}]{jaloveckas2023efficient}%
  \BibitemOpen
  \bibfield  {author} {\bibinfo {author} {\bibfnamefont {J.~E.}\ \bibnamefont {Jaloveckas}}, \bibinfo {author} {\bibfnamefont {M.~T.~P.}\ \bibnamefont {Nguyen}}, \bibinfo {author} {\bibfnamefont {L.}~\bibnamefont {Palackal}}, \bibinfo {author} {\bibfnamefont {J.~M.}\ \bibnamefont {Lorenz}},\ and\ \bibinfo {author} {\bibfnamefont {H.}~\bibnamefont {Ehm}},\ }\href@noop {} {\bibinfo {title} {Efficient learning of {S}parse {P}auli {L}indblad models for fully connected qubit topology}} (\bibinfo {year} {2023}),\ \Eprint {https://arxiv.org/abs/2311.11639} {arXiv:2311.11639 [quant-ph]} \BibitemShut {NoStop}%
\bibitem [{\citenamefont {van~den Berg}\ and\ \citenamefont {Wocjan}(2024)}]{berg2024techniques}%
  \BibitemOpen
  \bibfield  {author} {\bibinfo {author} {\bibfnamefont {E.}~\bibnamefont {van~den Berg}}\ and\ \bibinfo {author} {\bibfnamefont {P.}~\bibnamefont {Wocjan}},\ }\href {https://doi.org/10.22331/q-2024-12-10-1556} {\bibfield  {journal} {\bibinfo  {journal} {{Quantum}}\ }\textbf {\bibinfo {volume} {8}},\ \bibinfo {pages} {1556} (\bibinfo {year} {2024})}\BibitemShut {NoStop}%
\bibitem [{\citenamefont {Hansenne}\ \emph {et~al.}(2025)\citenamefont {Hansenne}, \citenamefont {Qu}, \citenamefont {Weinbrenner}, \citenamefont {de~Gois}, \citenamefont {Wang} \emph {et~al.}}]{hansenne2024optimal}%
  \BibitemOpen
  \bibfield  {author} {\bibinfo {author} {\bibfnamefont {K.}~\bibnamefont {Hansenne}}, \bibinfo {author} {\bibfnamefont {R.}~\bibnamefont {Qu}}, \bibinfo {author} {\bibfnamefont {L.~T.}\ \bibnamefont {Weinbrenner}}, \bibinfo {author} {\bibfnamefont {C.}~\bibnamefont {de~Gois}}, \bibinfo {author} {\bibfnamefont {H.}~\bibnamefont {Wang}}, \emph {et~al.},\ }\href {https://doi.org/10.1103/t6qb-kdcp} {\bibfield  {journal} {\bibinfo  {journal} {Phys. Rev. Lett.}\ }\textbf {\bibinfo {volume} {135}},\ \bibinfo {pages} {060801} (\bibinfo {year} {2025})}\BibitemShut {NoStop}%
\bibitem [{\citenamefont {Veltheim}\ and\ \citenamefont {Keski-Vakkuri}(2025)}]{veltheim2024multiset}%
  \BibitemOpen
  \bibfield  {author} {\bibinfo {author} {\bibfnamefont {O.}~\bibnamefont {Veltheim}}\ and\ \bibinfo {author} {\bibfnamefont {E.}~\bibnamefont {Keski-Vakkuri}},\ }\href {https://doi.org/10.1103/PhysRevLett.134.030801} {\bibfield  {journal} {\bibinfo  {journal} {Phys. Rev. Lett.}\ }\textbf {\bibinfo {volume} {134}},\ \bibinfo {pages} {030801} (\bibinfo {year} {2025})}\BibitemShut {NoStop}%
\bibitem [{\citenamefont {Altland}\ and\ \citenamefont {Simons}(2010)}]{Altland_Simons_2010}%
  \BibitemOpen
  \bibfield  {author} {\bibinfo {author} {\bibfnamefont {A.}~\bibnamefont {Altland}}\ and\ \bibinfo {author} {\bibfnamefont {B.~D.}\ \bibnamefont {Simons}},\ }\href@noop {} {\emph {\bibinfo {title} {Condensed Matter Field Theory}}},\ \bibinfo {edition} {2nd}\ ed.\ (\bibinfo  {publisher} {Cambridge University Press},\ \bibinfo {year} {2010})\BibitemShut {NoStop}%
\bibitem [{\citenamefont {Vizing}(1964)}]{Vizing1964}%
  \BibitemOpen
  \bibfield  {author} {\bibinfo {author} {\bibfnamefont {V.~G.}\ \bibnamefont {Vizing}},\ }\href {https://cir.nii.ac.jp/crid/1571980075458819456} {\bibfield  {journal} {\bibinfo  {journal} {Discret Analiz}\ }\textbf {\bibinfo {volume} {3}},\ \bibinfo {pages} {25} (\bibinfo {year} {1964})}\BibitemShut {NoStop}%
\bibitem [{\citenamefont {Gupta}(1968)}]{gupta1968studies}%
  \BibitemOpen
  \bibfield  {author} {\bibinfo {author} {\bibfnamefont {R.~P.}\ \bibnamefont {Gupta}},\ }\emph {\bibinfo {title} {Studies in the Theory of Graphs}},\ \href@noop {} {\bibinfo {type} {Doctoral thesis}},\ \bibinfo  {school} {Indian Statistical Institute}, \bibinfo {address} {Kolkata} (\bibinfo {year} {1968})\BibitemShut {NoStop}%
\bibitem [{Note1()}]{Note1}%
  \BibitemOpen
  \bibinfo {note} {In general, the problem of determining whether $\chi ^\prime $ is $\Delta $ or $\Delta + 1$ is NP-complete \cite {Holyer1981,Garey1983}. On the other hand, as the optimality gap is at most one, not much is lost by taking the upper bound, which would incur at most two additional measurement settings. Crucially, in that case it is also easy to actually construct a colouring of this degree, and therefore to find which measurements to perform. This can be done for any graph in time complexity $O(\abs {V} \abs {E})$ via the Misra-Gries construction \cite {MISRAGRIES}, which can be improved for particular classes of graphs (see e.g.\ \cite [Table 1]{Cole2008} for an overview of edge colouring algorithms for planar graphs).}\BibitemShut {Stop}%
\bibitem [{\citenamefont {L\"owdin}(1955)}]{loewdin1955quantum}%
  \BibitemOpen
  \bibfield  {author} {\bibinfo {author} {\bibfnamefont {P.-O.}\ \bibnamefont {L\"owdin}},\ }\href {https://doi.org/10.1103/PhysRev.97.1474} {\bibfield  {journal} {\bibinfo  {journal} {Phys. Rev.}\ }\textbf {\bibinfo {volume} {97}},\ \bibinfo {pages} {1474} (\bibinfo {year} {1955})}\BibitemShut {NoStop}%
\bibitem [{\citenamefont {Coleman}(1963)}]{Coleman1963}%
  \BibitemOpen
  \bibfield  {author} {\bibinfo {author} {\bibfnamefont {A.~J.}\ \bibnamefont {Coleman}},\ }\href {https://doi.org/10.1103/RevModPhys.35.668} {\bibfield  {journal} {\bibinfo  {journal} {Rev. Mod. Phys.}\ }\textbf {\bibinfo {volume} {35}},\ \bibinfo {pages} {668} (\bibinfo {year} {1963})}\BibitemShut {NoStop}%
\bibitem [{\citenamefont {Orlin}(1977)}]{orlin1977contentment}%
  \BibitemOpen
  \bibfield  {author} {\bibinfo {author} {\bibfnamefont {J.}~\bibnamefont {Orlin}},\ }in\ \href {https://doi.org/10.1016/1385-7258(77)90055-5} {\emph {\bibinfo {booktitle} {Indagationes Mathematicae (Proceedings)}}},\ Vol.~\bibinfo {volume} {80}\ (\bibinfo {organization} {Elsevier},\ \bibinfo {year} {1977})\ pp.\ \bibinfo {pages} {406--424}\BibitemShut {NoStop}%
\bibitem [{\citenamefont {Kou}\ \emph {et~al.}(1978)\citenamefont {Kou}, \citenamefont {Stockmeyer},\ and\ \citenamefont {Wong}}]{kou1978covering}%
  \BibitemOpen
  \bibfield  {author} {\bibinfo {author} {\bibfnamefont {L.~T.}\ \bibnamefont {Kou}}, \bibinfo {author} {\bibfnamefont {L.~J.}\ \bibnamefont {Stockmeyer}},\ and\ \bibinfo {author} {\bibfnamefont {C.~K.}\ \bibnamefont {Wong}},\ }\href {https://doi.org/10.1145/359340.359346} {\bibfield  {journal} {\bibinfo  {journal} {Commun. ACM}\ }\textbf {\bibinfo {volume} {21}},\ \bibinfo {pages} {135–139} (\bibinfo {year} {1978})}\BibitemShut {NoStop}%
\bibitem [{\citenamefont {Hartman}(2005)}]{Hartman2005}%
  \BibitemOpen
  \bibfield  {author} {\bibinfo {author} {\bibfnamefont {A.}~\bibnamefont {Hartman}},\ }\bibinfo {title} {Software and hardware testing using combinatorial covering suites},\ in\ \href {https://doi.org/10.1007/0-387-25036-0_10} {\emph {\bibinfo {booktitle} {Graph Theory, Combinatorics and Algorithms: Interdisciplinary Applications}}},\ \bibinfo {editor} {edited by\ \bibinfo {editor} {\bibfnamefont {M.~C.}\ \bibnamefont {Golumbic}}\ and\ \bibinfo {editor} {\bibfnamefont {I.~B.-A.}\ \bibnamefont {Hartman}}}\ (\bibinfo  {publisher} {Springer US},\ \bibinfo {address} {Boston, MA},\ \bibinfo {year} {2005})\ pp.\ \bibinfo {pages} {237--266}\BibitemShut {NoStop}%
\bibitem [{\citenamefont {Torres-Jimenez}\ and\ \citenamefont {Izquierdo-Marquez}(2013)}]{torres2013survey}%
  \BibitemOpen
  \bibfield  {author} {\bibinfo {author} {\bibfnamefont {J.}~\bibnamefont {Torres-Jimenez}}\ and\ \bibinfo {author} {\bibfnamefont {I.}~\bibnamefont {Izquierdo-Marquez}},\ }in\ \href {https://doi.org/10.1109/SYNASC.2013.10} {\emph {\bibinfo {booktitle} {Proceedings of the 15th International Symposium on Symbolic and Numeric Algorithms for Scientific Computing}}}\ (\bibinfo {organization} {IEEE},\ \bibinfo {year} {2013})\ pp.\ \bibinfo {pages} {20--27}\BibitemShut {NoStop}%
\bibitem [{\citenamefont {Gramm}\ \emph {et~al.}()\citenamefont {Gramm}, \citenamefont {Guo}, \citenamefont {Hüffner},\ and\ \citenamefont {Niedermeier}}]{gramm2006datareduction}%
  \BibitemOpen
  \bibfield  {author} {\bibinfo {author} {\bibfnamefont {J.}~\bibnamefont {Gramm}}, \bibinfo {author} {\bibfnamefont {J.}~\bibnamefont {Guo}}, \bibinfo {author} {\bibfnamefont {F.}~\bibnamefont {Hüffner}},\ and\ \bibinfo {author} {\bibfnamefont {R.}~\bibnamefont {Niedermeier}},\ }\bibinfo {title} {Data reduction, exact, and heuristic algorithms for clique cover},\ in\ \href {https://doi.org/10.1137/1.9781611972863.9} {\emph {\bibinfo {booktitle} {2006 Proceedings of the Workshop on Algorithm Engineering and Experiments (ALENEX)}}},\ pp.\ \bibinfo {pages} {86--94}\BibitemShut {NoStop}%
\bibitem [{\citenamefont {Conte}\ \emph {et~al.}(2016)\citenamefont {Conte}, \citenamefont {Grossi},\ and\ \citenamefont {Marino}}]{conte_clique_2016}%
  \BibitemOpen
  \bibfield  {author} {\bibinfo {author} {\bibfnamefont {A.}~\bibnamefont {Conte}}, \bibinfo {author} {\bibfnamefont {R.}~\bibnamefont {Grossi}},\ and\ \bibinfo {author} {\bibfnamefont {A.}~\bibnamefont {Marino}},\ }in\ \href {https://doi.org/10.1145/2851613.2851816} {\emph {\bibinfo {booktitle} {Proceedings of the 31st {Annual} {ACM} {Symposium} on {Applied} {Computing}}}}\ (\bibinfo  {publisher} {ACM},\ \bibinfo {address} {Pisa, Italy},\ \bibinfo {year} {2016})\ pp.\ \bibinfo {pages} {1134--1139}\BibitemShut {NoStop}%
\bibitem [{\citenamefont {Wang}\ \emph {et~al.}(2020)\citenamefont {Wang}, \citenamefont {Hu}, \citenamefont {Sanders},\ and\ \citenamefont {Kais}}]{Wang_2020}%
  \BibitemOpen
  \bibfield  {author} {\bibinfo {author} {\bibfnamefont {Y.}~\bibnamefont {Wang}}, \bibinfo {author} {\bibfnamefont {Z.}~\bibnamefont {Hu}}, \bibinfo {author} {\bibfnamefont {B.~C.}\ \bibnamefont {Sanders}},\ and\ \bibinfo {author} {\bibfnamefont {S.}~\bibnamefont {Kais}},\ }\href {https://doi.org/10.3389/fphy.2020.589504} {\bibfield  {journal} {\bibinfo  {journal} {Frontiers in Physics}\ }\textbf {\bibinfo {volume} {8}},\ \bibinfo {pages} {589504} (\bibinfo {year} {2020})}\BibitemShut {NoStop}%
\bibitem [{\citenamefont {Colbourn}\ and\ \citenamefont {Dinitz}(2006)}]{handbook}%
  \BibitemOpen
  \bibfield  {author} {\bibinfo {author} {\bibfnamefont {C.}~\bibnamefont {Colbourn}}\ and\ \bibinfo {author} {\bibfnamefont {J.}~\bibnamefont {Dinitz}},\ }\href {https://doi.org/10.1201/9781420010541} {\emph {\bibinfo {title} {Handbook of Combinatorial Designs}}}\ (\bibinfo  {publisher} {Chapman and Hall/CRC},\ \bibinfo {year} {2006})\BibitemShut {NoStop}%
\bibitem [{\citenamefont {{OEIS Foundation Inc. (2024)}}()}]{oeisA000438}%
  \BibitemOpen
  \bibfield  {author} {\bibinfo {author} {\bibnamefont {{OEIS Foundation Inc. (2024)}}},\ }\href@noop {} {\bibinfo {title} {Entry {A000438} in {T}he {O}n-{L}ine {E}ncyclopedia of {I}nteger {S}equences}},\ \bibinfo {note} {\url{https://oeis.org/A000438}}\BibitemShut {NoStop}%
\bibitem [{\citenamefont {Holyer}(1981)}]{Holyer1981}%
  \BibitemOpen
  \bibfield  {author} {\bibinfo {author} {\bibfnamefont {I.}~\bibnamefont {Holyer}},\ }\href {https://doi.org/10.1137/0210054} {\bibfield  {journal} {\bibinfo  {journal} {SIAM J. Comput.}\ }\textbf {\bibinfo {volume} {10}},\ \bibinfo {pages} {713} (\bibinfo {year} {1981})}\BibitemShut {NoStop}%
\bibitem [{\citenamefont {Garey}\ and\ \citenamefont {Johnson}(1983)}]{Garey1983}%
  \BibitemOpen
  \bibfield  {author} {\bibinfo {author} {\bibfnamefont {M.~R.}\ \bibnamefont {Garey}}\ and\ \bibinfo {author} {\bibfnamefont {D.~S.}\ \bibnamefont {Johnson}},\ }\href {https://doi.org/10.2307/2273574} {\bibfield  {journal} {\bibinfo  {journal} {J. Symb. Log.}\ }\textbf {\bibinfo {volume} {48}},\ \bibinfo {pages} {498} (\bibinfo {year} {1983})}\BibitemShut {NoStop}%
\bibitem [{\citenamefont {Misra}\ and\ \citenamefont {Gries}(1992)}]{MISRAGRIES}%
  \BibitemOpen
  \bibfield  {author} {\bibinfo {author} {\bibfnamefont {J.}~\bibnamefont {Misra}}\ and\ \bibinfo {author} {\bibfnamefont {D.}~\bibnamefont {Gries}},\ }\href {https://doi.org/https://doi.org/10.1016/0020-0190(92)90041-S} {\bibfield  {journal} {\bibinfo  {journal} {Inf. Process. Lett.}\ }\textbf {\bibinfo {volume} {41}},\ \bibinfo {pages} {131} (\bibinfo {year} {1992})}\BibitemShut {NoStop}%
\bibitem [{\citenamefont {Cole}\ and\ \citenamefont {Kowalik}(2008)}]{Cole2008}%
  \BibitemOpen
  \bibfield  {author} {\bibinfo {author} {\bibfnamefont {R.}~\bibnamefont {Cole}}\ and\ \bibinfo {author} {\bibfnamefont {{\L}.}~\bibnamefont {Kowalik}},\ }\href {https://doi.org/10.1007/s00453-007-9044-3} {\bibfield  {journal} {\bibinfo  {journal} {Algorithmica}\ }\textbf {\bibinfo {volume} {50}},\ \bibinfo {pages} {351} (\bibinfo {year} {2008})}\BibitemShut {NoStop}%
\end{thebibliography}%

\end{document}